\pdfoutput=1
\documentclass[aps,prx,reprint,showpacs,superscriptaddress,notitlepage,floatfix]{revtex4-2}

\usepackage[pdftex]{graphicx}
\usepackage[pdftex]{epsfig}
\usepackage{epstopdf}
\usepackage{color}
\usepackage{amssymb,amsmath}
\usepackage{graphicx}
\usepackage{dcolumn}
\usepackage{multirow}
\usepackage{hyperref}
\usepackage{siunitx}

\usepackage{mathtools}
\DeclarePairedDelimiter\floor{\lfloor}{\rfloor}

\usepackage{enumitem,amssymb}
\newlist{todolist}{itemize}{2}
\setlist[todolist]{label=$\square$}

\begin{document}

%\cleardoublepage

\title{The impact of stochastic incorporation on atomic-precision Si:P arrays}
\author{Jeffrey A. Ivie}
\thanks{These authors contributed equally }
\affiliation{Sandia National Laboratories, Albuquerque NM, USA}
\author{Quinn Campbell}
\thanks{These authors contributed equally }
\affiliation{Center for Computing Research, Sandia National Laboratories, Albuquerque NM, USA}
\author{Justin C. Koepke}
\thanks{These authors contributed equally }
\affiliation{Sandia National Laboratories, Albuquerque NM, USA}
\author{Mitchell I. Brickson}
\affiliation{Center for Computing Research, Sandia National Laboratories, Albuquerque NM, USA}
\affiliation{Center for Quantum Information and Control (CQuIC), University of New Mexico, Albuquerque NM, USA}
\author{Peter A. Schultz}
\author{Richard P. Muller}
\affiliation{Sandia National Laboratories, Albuquerque NM, USA}
\author{Andrew M. Mounce}
\affiliation{Center for Integrated Nanotechnologies (CINT), Albuquerque NM, USA}
\author{Daniel R. Ward}
\altaffiliation[Present Address: ]{HRL Laboratories, LLC, Malibu, CA 90265}
\author{Malcolm S. Carroll}
\altaffiliation[Present Address: ]{Princeton Plasma Physics Laboratory,Princeton, NJ 08543}
\affiliation{Sandia National Laboratories, Albuquerque NM, USA}
\author{Ezra Bussmann}
\affiliation{Sandia National Laboratories, Albuquerque NM, USA}
\author{Andrew D. Baczewski}
\email[Corresponding Author: ]{adbacze@sandia.gov}
\affiliation{Center for Computing Research, Sandia National Laboratories, Albuquerque NM, USA}
\affiliation{Center for Quantum Information and Control (CQuIC), University of New Mexico, Albuquerque NM, USA}
\author{Shashank Misra}
\affiliation{Sandia National Laboratories, Albuquerque NM, USA}

\begin{abstract}
Scanning tunneling microscope lithography can be used to create nanoelectronic devices in which dopant atoms are precisely positioned in a Si lattice within $\sim$\SI{1}{\nano\meter} of a target position.
This exquisite precision is promising for realizing various quantum technologies. 
However, a potentially impactful form of disorder is due to incorporation kinetics, in which the number of P atoms that incorporate into a single lithographic window is manifestly uncertain.
We present experimental results indicating that the likelihood of incorporating into an ideally written three-dimer single-donor window is $63 \pm 10\%$ for room-temperature dosing, and corroborate these results with a model for the incorporation kinetics.
Nevertheless, further analysis of this model suggests conditions that might raise the incorporation rate to near-deterministic levels.
We simulate bias spectroscopy on a chain of comparable dimensions to the array in our yield study, indicating that such an experiment may help confirm the inferred incorporation rate.
\end{abstract}

\maketitle

\section*{Introduction}
\label{sec:introduction}

\begin{center}
\begin{figure*}[ht]
    \includegraphics[width=\textwidth]{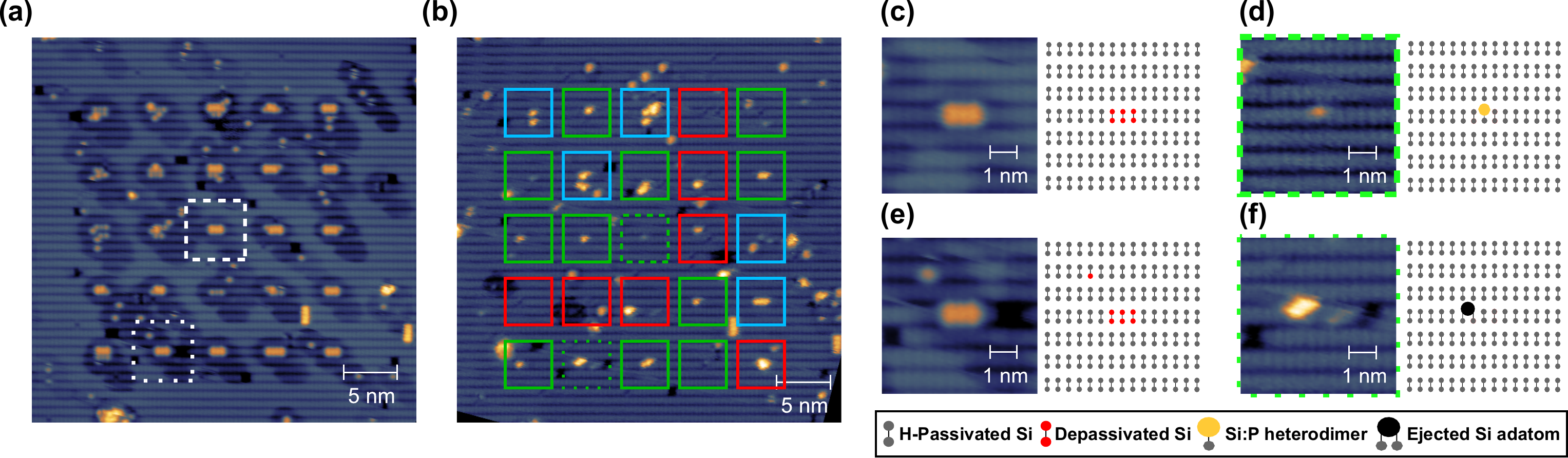}
    \caption{
    Example STM lithography array.
    (a) STM image of 25 patterned windows, where the two marked windows correspond to (c) and (e), respectively. 
    (b) STM image of the same windows after PH$_3$ dosing and incorporation. 
    Each window is colored to indicate the number of P incorporation features in each window, with red corresponding to 0, green corresponding to 1, and blue corresponding to 2. 
    The corresponding windows after incorporation for (c) and (e) are shown in (d) and (f), respectively.
    (c) Example $w=3$ window after hydrogen depassivation lithography corresponding to the middle window in (a). 
    (d) Illustration of the window from (c) after incorporation where a Si:P heterodimer remains. It is likely that the corresponding Si adatom has diffused away.
    (e) Another example $w=3$ window after lithography, corresponding to the window in the lower left in (a). 
    (f) Illustration of the window from (e) after incorporation where an ejected Si adatom is observed. This adatom feature is artificially broadened by a scan artifact and is likely masking any corresponding Si:P heterodimer feature.
    }
    \label{fig:overview}
\end{figure*}
\end{center}

Atomic precision (AP) placement of individual dopant atoms in Si nanoelectronic devices is a promising avenue for realizing a variety of technologies ranging from analog quantum simulators~\cite{georgescu2014quantum,prati2012anderson,prati2016band,le2017extended,dusko2018adequacy,altman2021quantum}, to qubits~\cite{buch2013spin,hill2015surface,pakkiam2018characterization,pakkiam2018single,he2019two,koch2019spin,bussmann2021apam}, to digital electronics~\cite{vskerevn2018cmos,ward2020atomic}.
This paper considers a particular limitation of AP donor placement with scanning tunneling microscope (STM) lithography~\cite{lyding1994nanoscale,schofield2003placement,randall2009atomic} (see Fig.~\ref{fig:overview}).
STM lithography allows for the fabrication of devices in which single donor atoms are positioned to within $\sim$\SI{1}{\nano\meter} of a target lattice site~\cite{oberbeck2004measurement}.
This weak placement disorder has previously been predicted not to be of concern when it comes to the prospects for realizing analog quantum simulators~\cite{dusko2018adequacy} or nuclear spin qubits~\cite{gamble2015multivalley} with this approach.
However, these models do not account for the failure of a precisely placed donor to incorporate.
While STM lithography achieves AP placement of individual donor atoms, we will show that a single precisely placed donor will incorporate with probability less than one for room-temperature dosing.
While this is an indirect observation and quantitatively similar to previously reported results~\cite{Fuchsle2011thesis}, our inference is corroborated by a kinetic model.
That same model suggests conditions that might lead to deterministic incorporation.

Analog quantum simulation of the Fermi-Hubbard model using donor arrays provides a useful example that both introduces the physics of interest and the potential impacts of stochastic incorporation.
The physics that is most relevant to analog quantum simulation is the hydrogenic nature of shallow donors~\cite{ramdas1981spectroscopy}.
In its neutral ground state, a single bulk-like P donor will have one weakly bound electron \SI{45.59}{\milli \eV} from the bulk conduction band edge~\cite{morin1954impurity,jagannath1981linewidths,mayur1993redetermination}. 
The affiliated orbital will have an anisotropic effective Bohr radius that exceeds that of a hydrogen atom in vacuum~\cite{kohn1955theory}. 
This renormalization is due to the bulk dielectric constant of the host crystal ($\epsilon_{Si}=11.7$) and the anisotropic conduction band effective mass ($m_{\perp}=0.19 m_e$ and $m_{\parallel}=0.92 m_e$).
This artificial hydrogen atom can also be positively charged when it is stripped of its single weakly bound electron, or negatively charge when it is doubly occupied.

The low-energy effective theory that governs the behavior of the weakly bound electrons on a chain or array of such artificial hydrogen atoms is an extended Fermi-Hubbard model~\cite{le2017extended,dusko2018adequacy}.
The tunnel coupling between sites is determined by the physical distance between donor atoms and it is feasible to realize spacings that are within an order of magnitude of the effective Bohr radius.
This can then realize instances of the extended Fermi-Hubbard model that are ``hard'' in the sense that there is competition between itinerance and localization of the weakly bound electrons.

Atomically precise arrays of more than $\sim$50 donors are needed to realize a system that is large enough to represent an instance of the extended Fermi-Hubbard model that exceeds the scale for which exact simulation on a classical computer is feasible~\footnote{This is a crude boundary that is based on whether the entire state vector of the system could be stored in classical memory. 
Approximate classical simulation methods can certainly exceed this scale.}.
Prior theoretical analyses~\cite{le2017extended,dusko2018adequacy} have shown that the placement disorder realized in STM lithography is sufficiently weak to realize large arrays that exhibit the desired physics, assuming that each lithographic window yields exactly one P donor placed to within $\sim$\SI{1}{\nano\meter}.
While current fabrication processes are evidently capable of reaching this precision in placement, the ultimate limits to their yield remains an open question.

To model the yield of STM lithography, we consider the probability of successfully fabricating an array of $N$ sites in which each site consists of $n=1$ donor,
\begin{equation}
    P(N) = \left(\sum \limits_{w} P_L(w) P_I(n=1|w) \right)^N. \label{eq:probability}
\end{equation}
Here $P_L(w)$ is the probability of patterning a lithographic window consisting of $w$ dimers within a single row and $P_{I}(n|w)$ is the probability of incorporating $n$ donors into such a window.
Implicit in this formula is the independence of success probabilities from site to site and the neglect of failure modes outside of lithography and incorporation chemistry~\footnote{For example, we assume that the probability of the sample being misplaced or stolen by a grey jay can be made arbitrarily close to zero.}.
Both assumptions make our formula an optimistic estimate.
$P_L(w)$ is ultimately a function of the STM hardware and control software whereas $P_{I}(n|w)$ is a function of the surface chemistry of window $w$.
To reliably fabricate arrays with $N >> 1$ sites, i.e., $P(N) \approx 1$, it is evident that a window must be identified such that $P_{I}(n=1|w)^N \approx 1$.
The data in Fig.~\ref{fig:statistics} suggest that even for the best choice of $w$, a window consisting of 3 dimers, $P_{I}(n=1|w=3) = 63 \pm 10\%$.
Even with perfect lithography and processing the probability of successfully fabricating a $\sim$50-donor array for analog quantum simulation is $\sim$10$^{-10}$.

In what follows, we substantiate this dismal observation with a model of the incorporation chemistry.
However, we also suggest a path forward.
While it may at first seem as if only $P_L(w)$ can be driven arbitrarily close to 1 through improvements to the STM hardware and control software, our incorporation model indicates that dosing at lower pressures for longer times and/or slightly higher temperatures can lead to $P_{I}(n|w)\rightarrow 1$.
Intuitively, the $w=3$ windows need sufficient time and space for a single PH$_3$ molecule to land in the window without interference and then follow the correct pathway to a configuration in which incorporation occurs.
Our model suggests that the stochastic incorporation that we have observed is potentially a feature of room-temperature dosing.
Because our incorporation statistics are based on inference, we briefly consider more direct measurable signatures of missing donors.
We propose that bias spectroscopy measurements on a chain of dimensions comparable to the array fabricated in this work can yield insights into whether stochastic incorporation is occurring in a given process.

\section*{Results}
\label{sec:results}

\subsection*{STM experiments}
\label{subsec:stm_experiments_results}

To generate a statistical sample sufficient to estimate $P_I(n|w)$ for different values of $n$ and $w$, 50 windows were patterned using standard STM lithography techniques (see Fig.~\ref{fig:overview}).
Our aim was to pattern an array consisting entirely of $w=3$ dimer windows into which the incorporation of $n=1$ P atom should be maximized.
While feedback lithography would result in near-perfect patterning of any target window size~\cite{hersam2000isolating,achal2018lithography,Wyrick2019devices}, we did not make use of it. 
Thus $P_L(w=3)$ is less than one but this is irrelevant for our analysis as subsequent scans of the depassivated windows were taken to establish $w$ for each window.
After dosing and incorporation, STM scans of the areas containing the depassivated windows were taken again to determine $n$.
Estimates for $P_I(n|w)$ are the relative frequencies of each observed $n$ for the measured $w$.

\begin{center}
\begin{figure*}[ht]
    \centering
     \includegraphics[width=\textwidth]{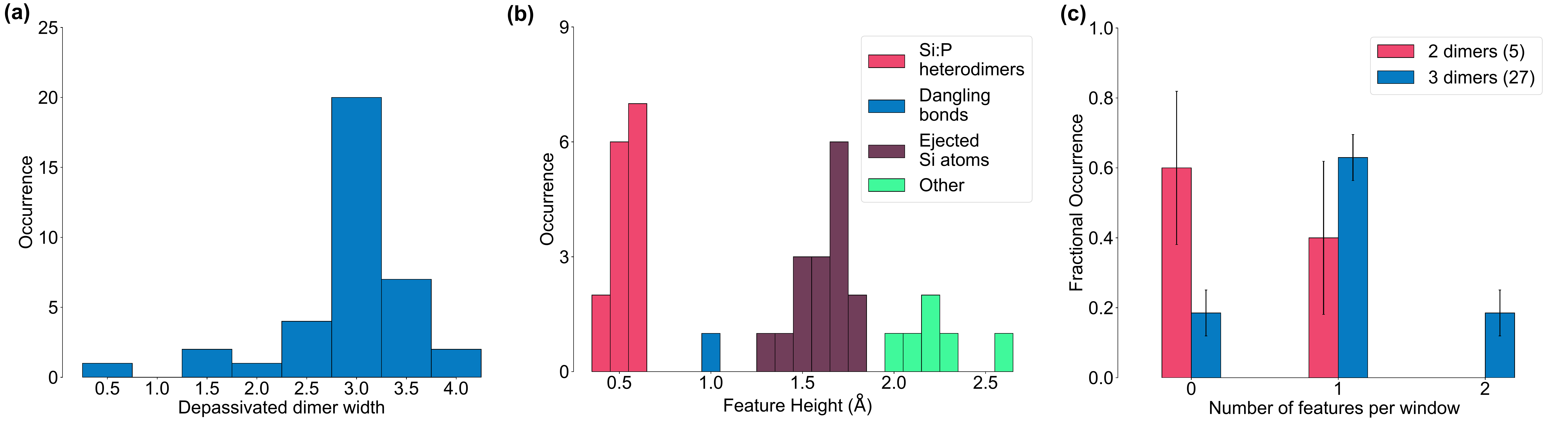}
    \caption{
    Statistics for our 50-window sample. 
    (a) The occurrence of different depassivation feature widths in the pre-dose images for a target width of $w=3$. These data are used to assign $w$ for a particular window.
    (b) The occurrence of distinct feature categories according to their height. These data are used to assign $n$ for a particular window.
    (c) The relative frequencies, $P_I(n|w)$ for $n=0,1,2$ and $w=2,3$. The number of windows realizing $w=2,3$ are indicated parenthetically.}
    \label{fig:statistics}
\end{figure*}
\end{center}

We note that STM scans of the pre-dose and post-incorporation surfaces were adjusted to bring the two scans into registration.
The post-incorporation scan is aligned to the pre-dose scan using intrinsic defects in the reconstructed Si$(100)$-$2\times$1 surface as markers (see Appendix B).
After alignment, each window in both the pre-dose and post-incorporation STM scans are extracted and gridded with respect to each Si lattice site (lattice constant 5.4 \AA) for analysis.
We expect that this alignment procedure results in a near-perfect overlap between the scans, where, in principle, only features present within a given window would be considered for evidence of incorporation. 
We allow for a single exception to this rule, where a Si:P heterodimer, as defined below, is allowed to be a single $2\times1$ dimer away from the window (Appendix B), as Si:P heterodimers represent direct evidence of P incorporation.
It should be noted that windows that demonstrated no change between STM scans of the post-dosing and post-incorporation surfaces are removed from the sample as incorporation cannot be established for such windows (total of 3).

%In order to generate the necessary statistics for the calculation of $P_I(n|w = 3)$, a total 50 different individual windows were patterning using standard AP patterning techniques (Methods), without the use of hydrogen feedback lithography (\jeff{citations…}. 
%While the use of feedback lithography would result in near-perfect patterning of the target window size ($w$), the large number of windows considered here allow for the generation of sufficient statistics for the determination of $P_I(n|w=3)$. 
%(\jeff{Jeff Note: I presume we will include the STM images in the SI})
%Due to the nature of the dosing and incorporation processes, 

The value of $w$ was determined for a given lithographic window by counting the number of contiguous depassivated Si atoms within a given dimer row. 
Depassivated atoms found in adjacent dimer rows or located non-contiguously to the primary pattern were excluded.
If no more than one depassivated Si atom satisfied the criteria set above, then the pattern was considered disordered and removed from our sample. 
The variability in the number of depassivated Si atoms for the 47 windows is depicted in Fig.~\ref{fig:statistics}a.
Here and elsewhere we specify $w$, the number of depassivated dimers rather than atoms, resulting in half-integer values for dimers that are partially depassivated.

For our target $w=3$ pattern, only 20 windows were determined to meet the above criteria, i.e., $P_L(w=3)=43\%$.
However, based on chemical considerations~\cite{warschkow2016reaction, wilson2006thermal} the $w=3.5$ pattern should exhibit similar incorporation chemistry, and including these windows in our sample increases the number of windows that realize the target pattern to 27, i.e., $P_L(w=3)=57\%$.
Thus the estimates of $P_I(n|w)$ for $w=2, 3$ in Fig.~\ref{fig:statistics} include the corresponding half-dimer results (i.e., $w \rightarrow \floor{w}$, see Appendix A for results separating the half-integer values of $w$).

To determine the number of donors in a given window from the post-incorporation scans, it was necessary to categorize features according to their height relative to the surrounding H-resist. 
The distribution of feature heights in the $w=2,3$ windows are shown in Fig.~\ref{fig:statistics}.
For a single donor to have incorporated successfully, two features should be present~\cite{curson2004stm, wilson2006thermal}--- an incorporated P atom in the form of a Si:P heterodimer and an ejected Si adatom. 
Based on methodology used previously in the literature~\cite{Fuchsle2011thesis}, we assign the first feature height grouping to Si:P heterodimers, with a relative height of 0.3-0.6 \AA \cite{curson2004stm, reusch2006phosphorus,oberbeck2002encapsulation} and the second to Si adatoms, with a relative height of 1.2-1.8 \AA~\cite{Fuchsle2011thesis}.

However, we also observe two feature categories that aren't associated with successful incorporation.
The first is a set of features with a relative height of 1 \AA~that was previously assigned to depassivated Si atoms~\cite{wang1994direct}, consistent with height measurements of the pre-dose surface.
The second set of features have relative heights greater than 1.8 \AA, which we ascribe to either background contamination in the UHV chamber or ejected Si adatoms adsorbed on H \cite{nara2008formation,kajiyama2005room}. We presume that these possible Si adatoms are purely adventitious as they never correspond to windows with zero features, and therefore neglect them in further analysis.
%In order to determine the number of features within a window post-incorporation, establishment of the different features present within a window, considering the uncertainty in window definition as established above, is necessary. 
%Different features can be distinguished by the corresponding height relative to the surrounding H-resist, with the distribution of feature heights in the $w=2,3$ windows shown in Fig.~\ref{fig:statistics}b. 
%For a successful incorporation event, two features should be present\cite{wilson2006thermal, curson2004stm}: an incorporated P atom in the form of Si:P heterodimer and an ejected Si adatom. 
%Features with a relative height of 1 $\AA$ have previously been assigned to depassivated Si atoms \cite{wang1994direct}, consistent with height measurements of the pre-dosing surface.
%A small subset of features has relative heights greater than 1.8 $\AA$, and without further characterization beyond the scope of this manuscript, these features are prescribed to background contamination in the UHV chamber and will not be discussed further.

Having assigned values of $w$ to each window and categorized the features in the post-incorporation scans, we evaluated $P_I(n|w)$ from the relative frequencies of particular values of $n$.
This required us to assign values of $n$ to a particular window based upon the presence of particular features, specifically the Si:P heterodimer and Si adatom features.
We note that both features were counted individually in Fig.~\ref{fig:statistics}b.
Due to the different diffusion barriers for the Si:P heterodimer~\cite{bennett2009diffusion, reusch2006phosphorus} and the Si adatom~\cite{jeong1997adsorption, jeong1999complex}, we do not expect that both features are necessarily present within a particular lithographic window.
Thus, if \textit{either} feature was found within a particular window it was counted toward an incorporation event.
This brings us to one of our key results, that $P_I(n=1|w=3)=63 \pm 10\%$, consistent with previously reported statistics for similarly sized windows~\cite{Fuchsle2011thesis}.

\begin{figure*}[ht]
    \centering
    \includegraphics[width=\textwidth]{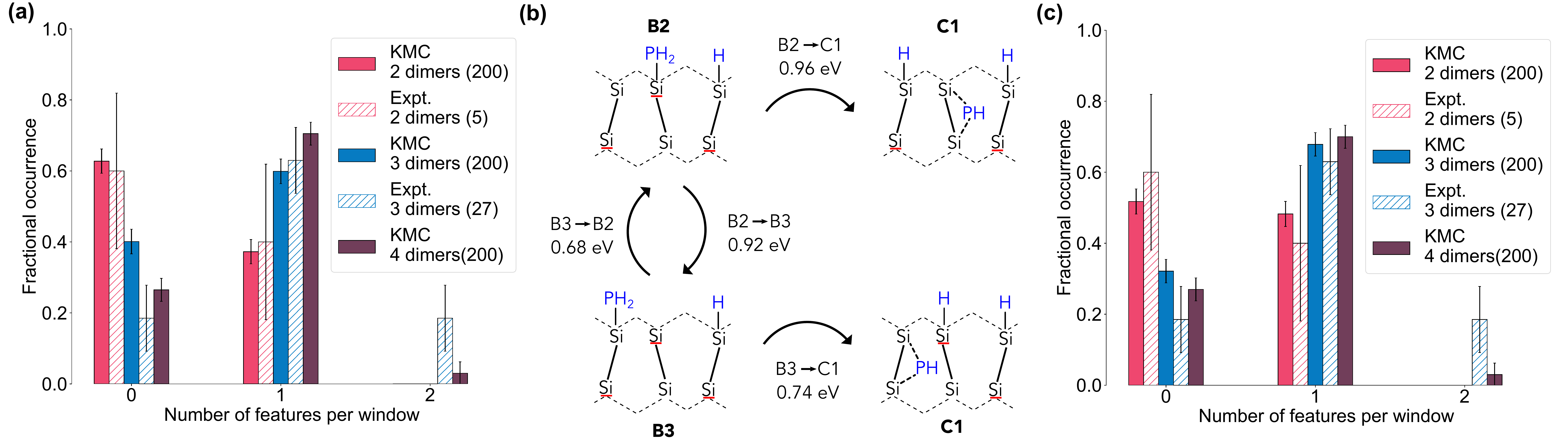}
    \caption{Comparison of our incorporation model with our experimental results.
    (a) $P_I(n|w)$ calculated using KMC versus experimentally measured data. 
(b) P likely incorporates via two different pathways: a one-step process, or a two-step process with a thermodynamically unfavorable middle step. 
We indicate the lower end of the dimer row by underlining the Si label in red. 
Labels here refer to the notation used by Warschkow \textit{et al.}~\cite{warschkow2016reaction}.
(c) Incorporated P calculated using an instance of our KMC model in which the barrier of the B3 to C1 reaction is lowered by \SI{0.1}{\eV}. 
Incorporation is generally increased, matching 3-dimer incorporation to the experimentally measured probability
\label{fig:KMC_results}}
\end{figure*}

We note that we expect $P_I(n>1|w=3)=0\%$ due to a paucity of dangling bonds on which 2 PH$_3$ molecules might shed 4 H atoms to arrive at a bridging P-H configuration in which the P can incorporate.
2-donor incorporation requires at least $w=4$ depassivated dimers.
However, we observe a small but non-zero probability for $P_I(n=2|w=3)$.
We hypothesize that this is a consequence of features diffusing into the window, rather than 2 donors incorporating within the same $w=3$ window.
We suggest that a similar mechanism might also be biasing our estimate of $P_I(n=1|w=2)$ upwards.
We note that this bias is \textit{not} accounted for in our quoted uncertainties.

%It should be noted that the presence of both a Si:P heterodimer and Si adatom feature in a window will only be counted as a single-donor incorporation event ($n=1$), however both features are noted in Fig.~\ref{fig:statistics}b. 
%However, due to the different diffusion barriers of the Si:P heterodimer \cite{bennett2009diffusion, reusch2006phosphorus} and the Si adatom \cite{jeong1997adsorption, jeong1999complex}, it is not expected that both features will be present within a lithographic window, hence justifying the identification of an incorporation feature for the assignment $n$ using only one of the expected features. 
%Appropriately sized windows ($w=3$) where two P atoms appear to have been incorporated ($n=2$) are hypothesized to be the consequence of appropriate features external to the window diffusing into the window, and not multiple donor incorporation events. 
%The same physical mechanism is proposed to have contributed to the increased $P_I(n=1|w=2)$ value for 2-dimer window data set, coupled with the sparsity of the data set.
%While $P_I(n=1|w=3)=68 \pm 10\%$ appears to be consistent with the current state-of-the-art dopant placement for single-donor sites, as noted earlier this is insufficient for constructing large chains or arrays consisting of single-donor sites. 
%Because only windows where $w = 3, 3.5$ are considered here, imperfections in the lithography can be ruled out as driving factor behind the low incorporation probability.

Returning to Eq.~\ref{eq:probability}, our results indicate that $P(N)=(40\% \pm 22\%)^N$ for our particular process~\footnote{This comes from $P(N) = (P_L(w=2) P_I(n=1|w=2)+ P_L(w=3) P_I(n=1|w=3))^N=(10\% \times(40\% \pm 20\%) + 57\% \times (63\% \pm 10\%))^N=(40\% \pm 22\%)^N$}.
%Returning to Eq.~\ref{eq:probability}, our results indicate that
%\begin{align}
%    P(N) = &(P_L(w=2) P_I(n=1|w=2) \nonumber\\
%    & + P_L(w=3) P_I(n=1|w=3))^N \nonumber\\ 
%    = & (10\% \times(40\% \pm 20\%) \nonumber\\
%    & + 54\% \times (63 \% \pm 10\%))^N \nonumber\\
%    = &(38\% \pm 22\%)^N .
%\end{align}
%$P(N)= (P_L(w=2) P_I(n=1|w=2) + P_L(w=3) P_I(n=1|w=3))^N= (10\% \times(40\% \pm 20\%) + 54\% \times (63 \% \pm 10\%))^N=(38\% \pm 22\%)^N$.
However, if $P_L(w=3)\rightarrow 100\%$ using, e.g., feedback lithography, we see that $P(N) \rightarrow (63\% \pm 10\%)^N$.
In both cases, the prospects for incorporating a large array for analog quantum simulation are dismal.
%That $P(N)$ appears to get worse as we get better at writing the desired $w=3$ pattern is an artifact of our identification scheme tending to bias incorporation rates upwards and our small sample size (i.e., there were 7 windows for which $w=2$).
To better understand these statistics we corroborate our results with an atomistic model that accounts for stochastic single-donor incorporation.

\subsection*{Incorporation model}
\label{subsec:incorporation_model}

We developed a Kinetic Monte Carlo (KMC) model to estimate $P_I(n|w)$.
Our model includes reaction barriers from Warschkow \textit{et al.}~\cite{warschkow2016reaction} and Wilson {\it et al.}~\cite{wilson2006thermal} with additional features described in the supplementary information.
Focusing particularly on the windows for which we expect single-donor incorporation, we compute $P_I(n|w)$ for $w=2, 3,$ and $4$.
The predictions of our model are compared to our experimental results in Fig.~\ref{fig:KMC_results}a. 
We match the ratio of half dimers included in each width bin to the measured ratio from this work.

For $n=1$ and $w=2,3$ our KMC model matches the experimental incorporation rate within error bars.
In the three dimer wide window, single-donor case, in particular, we predict $P_I(n=1|w=3)=59\% \pm 2\%$ compared to the measured value of $63\% \pm 10\%$.
For each reaction within our model chemistry, the rate is governed by an attempt frequency and a reaction barrier according to the Arrhenius equation.
As the rates are exponentially sensitive to the reaction barriers and errors $\sim$\SI{0.1}{\eV} are typical of the semilocal DFT calculations that predicted them, we examine whether we can better reproduce experiment by comparably sized adjustments to our barrier heights.
Rather than assessing the sensitivity of incorporation rates to every barrier independently, we first identify particularly sensitive steps in the reaction pathway.

The two reaction pathways that dominate the incorporation of a P atom are depicted in Fig.~\ref{fig:KMC_results}b. 
In the first pathway, outlined in the supporting information, a PH$_2$ molecule in a lower dimer end position (B2, using the naming conventions of Warschkow \textit{et al.}\cite{warschkow2016reaction}) moves into a bridging position between the two dimer atoms while simultaneously losing a hydrogen to the nearby dimer (C1). 
Alternatively, the PH$_2$ molecule can first move to a nearby dimer, but on the raised end of the dimer (B3), before finally moving to a bridging PH position (C1). 
This latter pathway involves a two-step process to incorporation, where the middle step is not thermodynamically favorable. 
Since the barrier for moving from B2 to C1 and the barrier for moving from B2 to B3 are within \SI{0.05}{\eV} of each other, a system that starts in a B2 configuration has a roughly equal chance of moving directly to C1 and incorporating or moving to B3. 
From B3, the system again has a roughly equal chance of incorporating to C1 or transitioning back to B2. 

This balance of options in the two-step pathway means that even slight changes that increase the likelihood of moving from B3 to C1 instead of from B3 back to B2 can have significant impacts on the final incorporation.
In Fig.~\ref{fig:KMC_results}c, we examine the impact of lowering the B3 to C1 barrier by \SI{0.1}{\eV} while keeping all other reaction barriers constant. 
We find that lowering the barrier for the B3 to C1 reaction leads to $P_I(n=1|w=3)=67\% \pm 2\%$. 
Further decreases to this reaction barrier eventually saturate before deterministic incorporation is achieved: lowering the B3 to C1 barrier by \SI{0.2}{\eV}, for instance, only results in $P_I(n=1|w=3)=70\% \pm 2\%$.

Accordingly, it is worth examining our model to see whether we can increase $P_{I}(n=1|w=3)$ to $100\%$.
We also consider whether we can corroborate the number of donors in an incorporated device through measurements other than the number of ejected Si adatoms and Si:P heterodimers.

\section*{Discussions and Conclusion}
\label{sec:discussions_and_conclusion}
\subsection*{Potential for deterministic incorporation}
\label{subsec:deterministic_incorporation_discussion}

\begin{figure*}[ht]
    \centering
    \includegraphics[width=\textwidth]{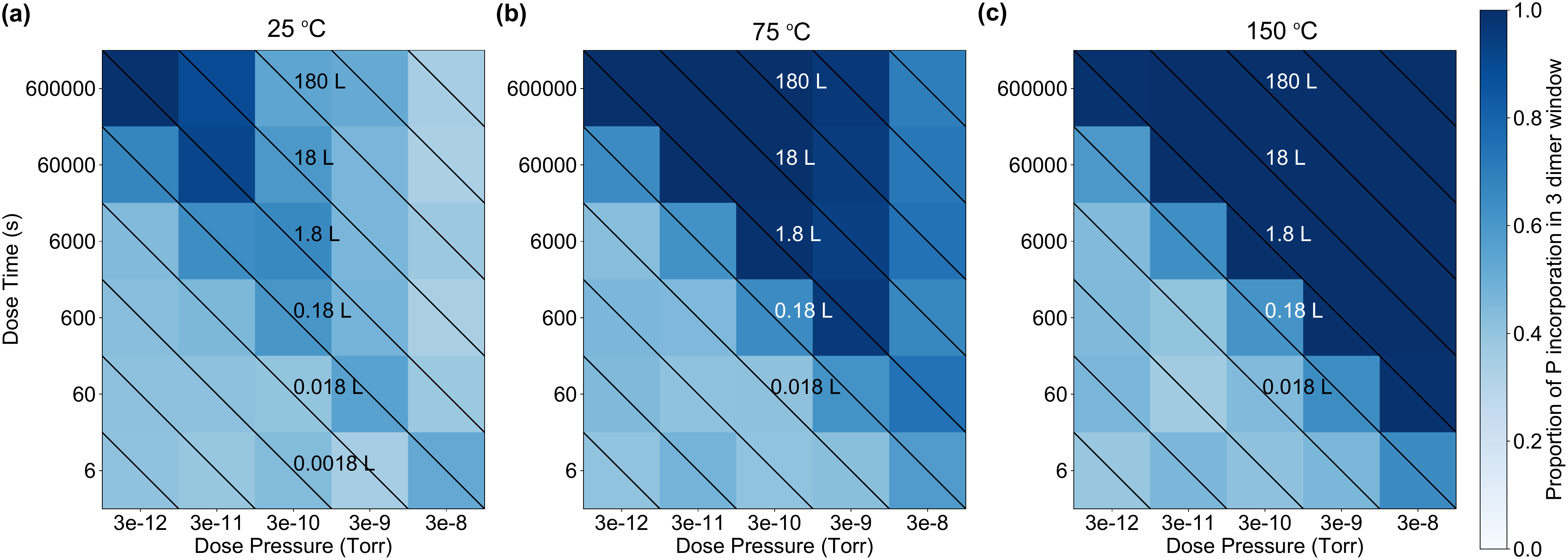}
    \caption{The simulated probability of incorporation as a function of both dose time and pressure.
    Diagonal lines represent lines of constant exposure, indicated in Langmuirs (L).
    Even within constant exposure, moving to low-pressure and long-time regimes can dramatically increase incorporation.
    Near-deterministic incorporation is predicted in the extreme low-pressure, long-time limit at room temperature.
    Incorporation can also be improved by raising the temperature during dosing as demonstrated by the heatmaps at (a) \SI{25}{\celsius}, (b) \SI{75}{\celsius}, and (c) \SI{150}{\celsius}.
\label{fig:dose_heatmap}}
\end{figure*}

Dose time and pressure can also strongly affect incorporation and we present predictions for the incorporation rate as a function of both in Fig.~\ref{fig:dose_heatmap}.
Notably the incorporation varies even at constant exposure, indicated as diagonal lines. 
A system at room temperature exposed to a 0.18 Langmuir dose of PH$_3$ at a pressure of \SI{3e-8}{Torr} for 6 seconds has $P_I(n=1|w=3)=52\% \pm 2\%$, while the same exposure at a pressure of \SI{3e-11}{Torr} for 6,000 seconds has $P_I(n=1|w=3)=64\% \pm 2\%$. 
In general, moving toward conditions of lower pressure and longer dose time leads to increased incorporation.

In fact, for extreme conditions of low pressure and long time, we predict near-deterministic incorporation.
At \SI{3e-12}{Torr} for \SI{6e5}{s} ($\sim$6.9 days), $P_I(n=1|w=3)=99\% \pm 1\%$. 
Currently these conditions may be prohibitively difficult to achieve, thanks to the need for extreme ultrahigh vacuum and exceptionally pure PH$_3$ dosing over several days, but it is instructive to understand the factors that lead our model to predict this.

Given enough space and time for a single PH$_3$ molecule in a window to dissociate before another PH$_3$ molecule adsorbs in the same space, incorporation \emph{will} occur.
The thermodynamic pathway for PH$_3$ dissociation is entirely downhill. 
There are no dead ends in the path that would prevent it from eventually incorporating, only wrong turns that can be reversed given sufficient attempts. 
As pressure decreases, the likelihood of a single PH$_3$ molecule adsorbing into a $w=3$ window fully dissociating before additional PH$_3$ adsorption increases.
At \SI{3e-12}{Torr}, the average time for an incorporation event is \SI{3.4e5}{s}. 
If only one PH$_3$ molecule adsorbs onto the $w=3$ window during the \SI{6e5}{s} dose, that molecule will be able to to shed its hydrogen and incorporate without any competition from neighboring molecules. 
In contrast, for a dose pressure of \SI{3e-8}{Torr}, increasing the dose time can decrease the likelihood of incorporation due to additional phosphine molecules adsorbing into the window before the molecule has fully dissociated. 
Once PH$_3$ evolves into a bridging PH configuration, however, the barriers for returning to a PH$_2$ state are so high that any recombination is unlikely, even in the extreme time scales considered here.
Moving from a C1 configuration to a B3 configuration requires overcoming a barrier of 1.86 eV, giving an average reaction time of \SI{1.8e19}{s} at room temperature and \SI{6.4e3}{s} at the anneal temperature of \SI{320}{\celsius}. 

Incorporation rates can therefore be increased by ensuring that the first PH$_3$ molecule adsorbed on the surface dissociates before another PH$_3$ molecule is adsorbed, blocking its pathway to incorporation. 
By increasing the sample temperature during dosing, the typical time for the dissociation reaction can be decreased significantly, as demonstrated in Fig.~\ref{fig:dose_heatmap}b-c.
At temperatures as low as \SI{75}{\celsius}, near-deterministic incorporation is possible within a more typical range of pressures and temperatures, with conditions of \SI{3e-10}{Torr} for \SI{6000}{s} ($\sim$1.6 hours), $P_I(n=1|w=3)=98.8\% \pm 1\%$. 
This trend continues to expand at higher temperatures with all exposures $>$ 1 L producing deterministic incorporation at \SI{150}{\celsius}.
Given the likelihood of hydrogen migration ruining the fidelity of the lithography window at higher temperatures, we expect that finding a feasible combination of pressure, time, and moderate temperature will produce the highest chance of success. 

Our simulations predict that stochastic incorporation of a P donor in a $w=3$ window is \textit{not} due to any inherent limitations of the PH$_3$ chemistry. 
It is instead due to the limitations of current room-temperature dosing techniques. 
This suggests that heating the sample during dosing and/or the development of ultra-pure chemical precursors and the ability to dose at pressures at or below \SI{3e-12}{Torr} may make deterministic donor incorporation with atomic precision feasible using a PH$_3$ chemistry.
Our results also suggest that with current technology and dose temperatures, incorporation can still be increased by moving to lower pressures and longer dose times. 
We additionally consider the benefits of low-temperature ($\sim$\SI{100}{\kelvin}) dosing for deterministic incorporation in the supplementary information.

\subsection*{Corroborating incorporation statistics through transport measurements}
\label{subsec:corroborating_incorporation_statistics}

In assessing whether stochastic incorporation occurs in a given fabrication process, it is critical to consider other data that could support or refute incorporation statistics derived from observed Si adatoms and Si:P heterodimers.
We have already noted that our inferred value of $P_I(n>1|w=3)>0\%$ is suspect and it is desirable to have a direct measurement of the electrically active substituted donors, more so because these are the technologically relevant elements.
Bias spectroscopy is thus a natural choice and a powerful characterization technique that can be used to directly measure the spectrum of current-carrying many-body states in a donor array.
We consider the properties of a small donor array that might be probed with this technique to corroborate our incorporation statistics.

\begin{figure}
    \centering
    \scalebox{.5}{\includegraphics[width=\textwidth]{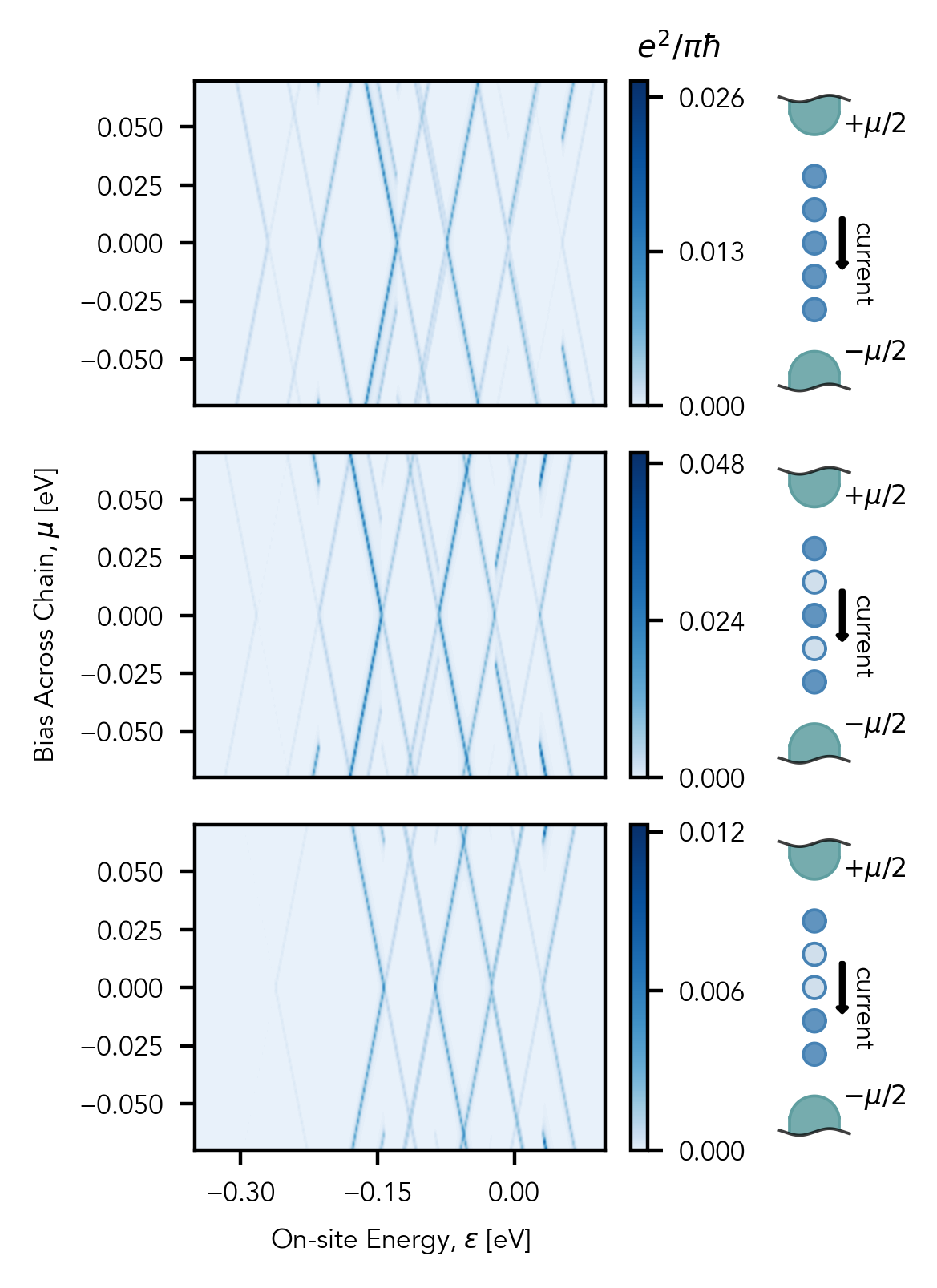}}
    \caption{Plots of the differential conductance (DC), $dI/d\mu$, as a function of the source-drain bias across the chain, $\mu$, and a uniform shift in the orbital energies, $\epsilon$, caused by a nearby gate voltage. 
    (a) DC across an ideal chain of 5 sites with 5 donors.
    (b) DC across a chain of 5 sites in which every other donor is missing. 
    (c) DC across a chain of 5 sites in which two consecutive donors are missing.
    In all cases, the faint transitions at higher occupancies (i.e., on the left of the plots) are diminished due to the increased impact of the intersite Coulomb interaction.
    We note that our calculations do not include current through scattering states to clearly demarcate the transport through the eigenstates of the chain and thus our Coulomb diamonds do not have signatures of transport ``above'' or ``below'' them, as would be expected in a real experiment.
    }
    \label{fig:simulated_arrays}
\end{figure}

A bias spectroscopy measurement would require fabricating such a donor array between two $\delta$-doped nanowire source and drain leads with either another in-plane lead or top gate~\cite{anderson2020low} to adjust the on-site potential across the array.
To reduce the likelihood of confounding the measurement with sites consisting of $n>1$ donors, we propose making an array comprised exclusively of $w =3$ windows.
Then it is likely the case that each site will consist of $n=0$ or $n=1$ donor.
A one-dimensional chain with a short spacing between sites will be ideal.
This spacing should be short enough that we can still expect conduction through the chain in case of missing sites.
The number of sites in the chain should be both short enough to maintain good stability in the patterning and long enough to have a reasonable expectation of missing donors.
A 5-site chain strikes a balance between being long enough to have 2 missing sites (in expectation) and short enough to expect good tip stability in patterning.
Indeed, a single row of the array in Fig.~\ref{fig:overview} would suffice.
For such a chain, the spacing between sites would correspond to a nearest-neighbor tunnel coupling of $\approx$ \SI{10}{\milli \eV}, diminishing to $\approx$ \SI{1}{\milli \eV} in the event of one missing site or $\approx$ \SI{0.1}{\milli \eV} for two consecutive missing sites~\footnote{These estimates are derived from predicted tunnel couplings between bulk-like donors taken from Ref.~\cite{gamble2015multivalley}.}.
Going to half this spacing is still viable and further increases the minimum tunnel coupling realized in the event of two missing sites by over an order of magnitude.
We consider simulations of such a chain in Fig.~\ref{fig:simulated_arrays}.

Our simulated bias spectra are intended to be qualitatively illustrative of the impacts of missing sites.
We present an exhaustive enumeration of all possible missing donors, as well as other spacings and orientations that may be of experimental interest in the supplementary information.
Here we simply compare the spectrum of the ideal chain to the most likely configurations consistent with $P_{I}(n=1|w=3) = 63 \pm 10\%$ (i.e., two missing donors).
We expect the spectrum for the chain in which every other donor is missing (1 of the 10 possible configurations for a five donor chain) to be qualitatively similar to that of the unbroken chain in this regime, with four prominent diamonds.
But the majority of configurations' bias spectra (9 of 10) will more closely resemble the case in which there are two consecutive donors and two missing sites, with three prominent diamonds, as demonstrated in the supporting information.
We thus expect that even in conditions favorable to tunnel coupling through missing donor sites, missing donors consistent with an incorporation rate of $\sim$60\% can be clearly detected with bias spectroscopy in 90\% of possible configurations for a five donor chain.
A larger spacing will yield even greater sensitivity to the precise locations of the missing donors.

A quantitatively accurate prediction of the current through a few-donor chain would require some improvements to our model that are discussed in the supplementary information.
However, we expect the basic principle illustrated in Fig.~\ref{fig:simulated_arrays} to hold--- for a sufficiently short chain there will still be measurable Coulomb diamonds even in the presence of missing donors and deviations from the transport signatures expected in the ideal chain might herald a particular pattern of missing donors.
We note that the transport signatures of missing donors are stronger for calculations using the traditional Fermi-Hubbard model (i.e., without the intersite Coulomb interaction).
In that limit, the incipient Hubbard bands above and below half filling give indications that clearly reflect the longest unbroken chain.
Because the intersite Coulomb interaction  in the extended Fermi-Hubbard model suppresses the formation of these incipient bands in small chains and breaks the symmetry between the bands above and below half-filling, we expect these signatures to be a bit messier.
Accordingly, we also expect that transport measurements of a chain of this dimension would provide experimental confirmation that the intersite term is a critical feature to include in effective models of Si:P arrays.

\subsection*{Conclusion}
\label{subsec:conclusion}

We demonstrated that even for perfectly patterned lithographic windows, single-donor incorporation is stochastic with a $63 \pm 10\%$ likelihood of success at room-temperature dosing conditions, which we corroborate with a kinetic model. 
We thus conclude that building classically intractable single donor arrays is overwhelmingly unlikely under these conditions.
Our model nevertheless suggests that near-deterministic incorporation may be possible above room temperature or at low-pressure, long-time dosing conditions.
We also considered bias spectroscopy as a method to directly measure the number of incorporated donors in an array, demonstrating that irregularities in the pattern of Coulomb diamonds can give clear indications of missing donors.

It is also worth remarking on the prospects of using AP technologies to realize analog quantum simulators with single acceptors~\cite{van2014probing,mol2015interface}, in addition to their prospective use in qubit technologies~\cite{ruskov2013chip,kobayashi2020engineering}.
The Fermi-Hubbard model parameters for a pair of substitutional B dopants have been measured in a sample prepared without AP placement~\cite{salfi2016quantum}.
With the recent demonstration of B$_2$H$_6$ as a precursor for delta doping~\cite{vskerevn2020bipolar}, it is clear that AP placement of individual B atoms might also be possible.
An incorporation model for B$_2$H$_6$, similar to the one in this paper, suggests that the tendency for dimerized B to incorporate as electrically inactive might limit the prospects for achieving deterministic single B atom placement for analog quantum simulation~\cite{campbell2021model}.
This is qualitatively different from the case presented here for PH$_3$, in which we predict that changing the pressure and temperature of the PH$_3$ dosing conditions can ultimately lead to deterministic AP placement.
Nevertheless, the model incorporation analysis also suggested that non-dimer acceptor precursors and halogen resists might provide a path forward~\cite{campbell2021model}.

Finally, it is worth emphasizing that stochastic incorporation is \emph{not} necessarily a no-go for the application of STM lithography to realizing analog quantum simulators, qubits, or digital electronics.
A simple workaround is to target the fabrication of multi-donor clusters~\cite{buch2013spin,hsueh2014spin} to realize quantum dots with precise dimensions.
While our results suggest that there will be uncertainty in the number of donors in such a cluster, it may be that there is a ``sweet spot'' in target cluster dimensions for which the resulting variance in the resulting application-relevant properties (e.g., tunnel couplings, charging energies, etc.) is sufficiently low.
In this context, the single-donor limit is somewhat pathological as there is a significant probability of \emph{not} fabricating a cluster at all.
The demonstration of a process that reliably realizes deterministic incorporation of a single donor into a three-dimer window, such as suggested by our modeling, would obviate some of the need for these considerations.
%Q note: Trying to make the ending a bit more snappy. A bit ambivalent on this specific change though...
%Our modeling suggests that this is possible and we hope to see this validated soon.

\section*{Methods}
\label{sec:methods}

\subsection*{STM experiments}
\label{subsec:stm_experiments}

Experiments were performed in a modified commercial UHV STM system (base pressure $<5 \times 10^{-10}$ Torr), plumbed directly with phosphine (PH$_3$, 99.9$\%$) and hydrogen (H$_2$, 99.999$\%$) gas lines. 
Si(100) samples (miscut $\le 0.1^{\circ}$), etched with appropriate alignment markers for pattern realignment, were prepared {\it ex-situ} by a wet chemical clean involving chemical oxidation (3:1 H$_2$SO$_4$:H$_2$O$_2$ 90 $^{\circ}$C, 5 mins), and reduction (10:1 H$_2$O:HF, 10 s), with rinsing with DI H$_2$O after each step, followed by successive sonications in MeOH, Acetone, IPA for 10 minutes. %Drawn from MBESTM_S53_H10. This information needs to have the alignment mark fabrication information added in some form - Done JAI%
After wet chemical cleaning, samples were dried with dry N$_2$ and immediately loaded into the UHV chamber. 
Samples were degassed at 625 $^{\circ}$C for 24 hours, then annealed for 10 s at 1100 $^{\circ}$C to achieve a clean Si(100) 2x1 reconstructed surface.  %Jeff - Strictly speaking, both are incorrect. However the degas is so complicated, it roughly comes out to 625 for 24 hours. Real degas is 400C for 1 hour, 645C for ~2 hours, 875 for 1 hour, ~625C for 24 hours. Degas from 12/6/2016 (initial degas). Sample was flashed... a lot so going to use final flash. Flash goes through ramps, with highest temperature being ~1100C for 10 sec. Flash from 12/20/2016
Temperatures were recorded with an IR pyrometer (Metis MP-25), with $50 ^{\circ}$C accuracy for temperatures $>600 ^{\circ}$C.

Samples were terminated with a monohydride resist for STM lithography by heating the sample to $350 ^{\circ}$C while exposing to atomic H supplied by cracking H$_2$ ($2 \times 10^{-6}$ Torr, 10 mins) with a W filament positioned 1 cm away from the sample. %Hydrogen termination pulled from 12/19/2106. Temperature is 342C -> 350C%
All STM operations utilized PtIr STM tips from NaugaNeedles (NN-USPtIr-W250, radius of curvature 25-50 nm), degassed at $\sim 175 ^{\circ}$C for 30 minutes and further heated by electron bombardment using a commercial Omicron tip preparation tool. %Tip is from Tip #59 notes, temperature of ~168C -> 175C. Estimating time since it was recorded for sure%
All AP patterns were generated using +3.5V bias, 6.0 nA tunneling current, 6.0 mC/cm dose, and 10 nm/sec tip speed. %pulled from 12/21/2016%

PH$_3$ is introduced into the UHV chamber via a precision leak valve, with PH$_3$ coverage calculated using the partial pressure of the PH$_3$ introduced during dosing measured by an RGA (specifically the 34 amu fragment). In a typical process, a PH$_3$ partial pressure of 3.0$\times $10$^{-10}$ Torr %from 12/21/2016 dose, thought this may be slightly off. But 0.15 vs 0.18L will not impact conclusions and we aren't redoing KMC calculations to fit a different value%
is introduced for 10 minutes at room temperature, resulting in a coverage of 0.15 Langmuir, which has been previously demonstrated to provide sufficient coverage for atomically precise donor structures~\cite{ward2017all,bussmann2015scanning}.
While the partial pressure reported through the PH$_3$ RGA value may represent a lower bound on the actual partial pressure introduced, the reported dosing conditions have also been demonstrated to provide the expected 0.25 ML coverage on a clean Si(100) surface for P incorporated at 310$ ^{\circ}$C \cite{lin2000interaction}, further demonstrating that the dosing conditions utilized will result in sufficient coverage in the depassivated windows.
A subsequent 310$ ^{\circ}$C anneal is used to incorporate P into Si lattice sites, demonstrated to be sufficient for full donor incorporation~\cite{mckibbin2009investigating}. %Drawn from 12/21/2016, actually 317C at max%

All STM scans shown were taken under -2.75V, 300 pA imaging conditions.

\subsection*{Theory and modeling}
\label{subsec:theory_and_modeling}

We use a Kinetic Monte Carlo model \cite{Bortz1975KMC,Gillespie1976KMC} as implemented in the \textsc{KMClib} package \cite{Leetmaa2014KMClib} to determine $P_I(n|w)$ as a function of $n$ and $w$.
Each KMC calculation is repeated 200 times with different random seeds, and the sample mean of the results is reported. 
We calculate error bars by assuming a binomial distribution of measured counts and using the standard error based on sample size. 
In all calculations unless otherwise stated, we use the experimental dose and anneal temperatures, pressures, and times. 

Our transport calculations are based on extended Fermi-Hubbard model parameters extracted from the multi-valley effective mass theory first presented in Ref.~\onlinecite{gamble2015multivalley}.
The Meir-Wingreen formula~\cite{meir1992landauer} is used to compute the current through interacting donor chains and the differential conductance is evaluated from the current using a finite difference formula.

\section*{Acknowledgements}
\label{sec:acknowledgements}

We gratefully acknowledge useful conversations with
Evan Anderson, % giving this a read
Josh Ballard,
Ed Bielejec, % meetings! 
DeAnna Campbell, % meetings and fab!
Udi Fuchs,
John Gamble, % goes without saying, at least in theory land
Toby Jacobson, % goes without saying, at least in theory land
Ehsan Khatami, % visited Shashank and Andrew during DQM
Dwight Luhman,
Michael Marshall,
Leon Maurer,
Andrea Morello, %
James Owen,
John Randall, % supporting conversations over the years?
Steve Rinaldi,
Scott Schmucker, % goes without saying - is co-authorship appropriate? I don't know how things work w/team experiment
Joe Simonson, % meetings!
Lisa Tracy, % meetings!
Rick Silver, % I'm sure we've all talked to Rick about this, at some point or another
and Esmeralda Yitamben. 
This work was partially supported by the Laboratory Directed Research and Development program at Sandia National Laboratories under projects 209242 (DQM), 213017 (FAIR DEAL), and 213048 (Quantum ACCESS). 
This work was also performed, in part, at the Center for Integrated Nanotechnologies, a U.S. DOE, Office of Basic Energy Sciences user facility.
Sandia National Laboratories is a multi-mission laboratory managed and operated by National Technology and Engineering Solutions of Sandia, LLC, a wholly owned subsidiary of Honeywell International, Inc., for DOE’s National Nuclear Security Administration under contract DE-NA0003525.

\section*{Author Contributions}
\label{sec:author_contributions}

JAI, QC, JCK contributed equally to this work.
EB, SM, and MSC coordinated the experimental effort.
ADB coordinated the theoretical effort and the writing of the manuscript.
JCK planned and conducted the experiments generating the STM data. 
DRW made critical advances to fabrication.
JAI, JCK, QC, and AMM analyzed the experimental data.
QC, PAS, and RPM developed the model for donor incorporation.
MIB carried out the transport calculations.

\section*{Competing Interests}

The authors declare no competing interests.

\bibliography{references}

% Supplemental Materials
% Supplementary information starts here
\clearpage
\widetext
\begin{center}
\textbf{\large Supplementary information: The impact of stochastic incorporation on atomic-precision Si:P arrays}
\end{center}

% Reset counters and add an S prefix on equation and figure numbers
\setcounter{section}{0}
\setcounter{page}{1}
\makeatletter

% This is Andrew's ongoing rewrite of the SM.
%%%%%%%%%%%%%%%%%%%%%%%%%%%%%%%%%%%%%%%%%%%%%%%%%%%%%%%%%%%%%%%%%%%%%%%%%%%%%%%%%%%%%%%%%%%%%%%%
\section*{Contents}

The supplementary information elaborates on details of some of the central results in the main body of the paper.
\begin{itemize}
    \item Appendix A describes experimental details regarding lithographic outcomes deviating from perfect two or three dimer windows and post-lithography/post-dose images.
    \item Appendix B describes the image registration procedure used to compare the pre-dose and post-incorporation STM scans. 
    \item Appendix C describes details of our incorporation model, including an analysis of a low-temperature process described in a recent patent (Ref.~\onlinecite{simmons_keizer_2019}).
    \item Appendix D describes the extended Fermi-Hubbard model and our simulated bias spectra on short donor chains.
\end{itemize}

\section*{Appendix A: Experimental details}
\setcounter{equation}{0}
\setcounter{figure}{0}
\renewcommand{\theequation}{A\arabic{equation}}
\renewcommand{\thefigure}{A\arabic{figure}}
\label{app:expt_details}

\begin{figure}[ht]
    \centering
    \includegraphics[width=\columnwidth]{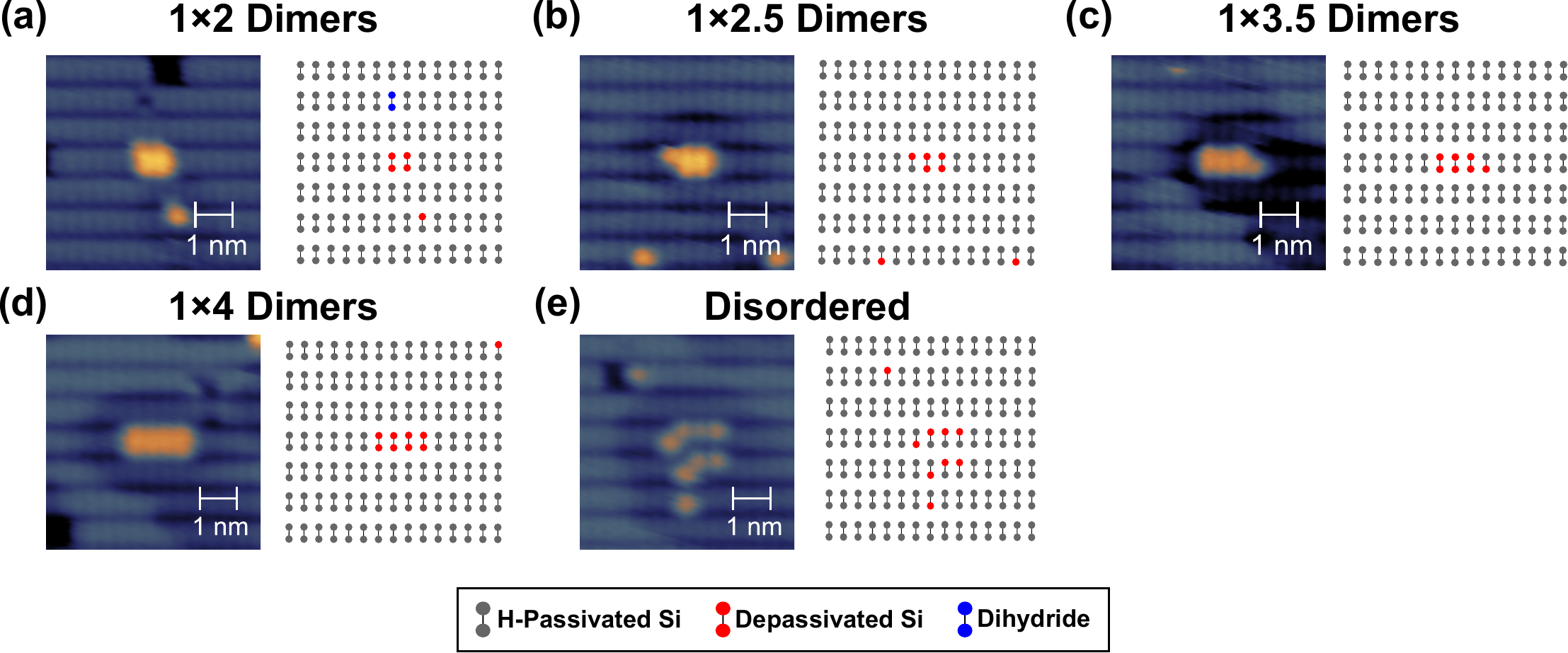}
    \caption{Common lithographic outcomes.}
    \label{fig:litho-outcomes}
\end{figure}

While using STM lithography to create three silicon dimer wide windows ($w=3$), we often create alternative outcomes such as windows of two ($w=2$) or four dimer ($w=4$) width, the inclusion of half dimers, and disordered arrays. 
We display a selection of common lithographic outcomes in Fig.~\ref{fig:litho-outcomes}.

After dosing the lithographic patterns with PH$_3$ gas, we examine the adsorption configurations of PH$_3$ fragments in the sites. 
We observe three general outcomes, as illustrated in Figure A2: (a) saturation of the lithographic template with PH$_3$ gas fragments, (b) less than saturation coverage of the lithographic template with PH$_3$ gas fragments, and infrequently (c) complete repassivation of the litho site with hydrogen. 
These observations of sites saturated with PH$_3$ fragments and sites with less than saturation coverage with PH$_3$ fragments occur in spite of dosing conditions which lead to ~0.25 ML P dopant density for large area patterns.

\begin{figure}[ht]
    \centering
    \includegraphics[width=\columnwidth]{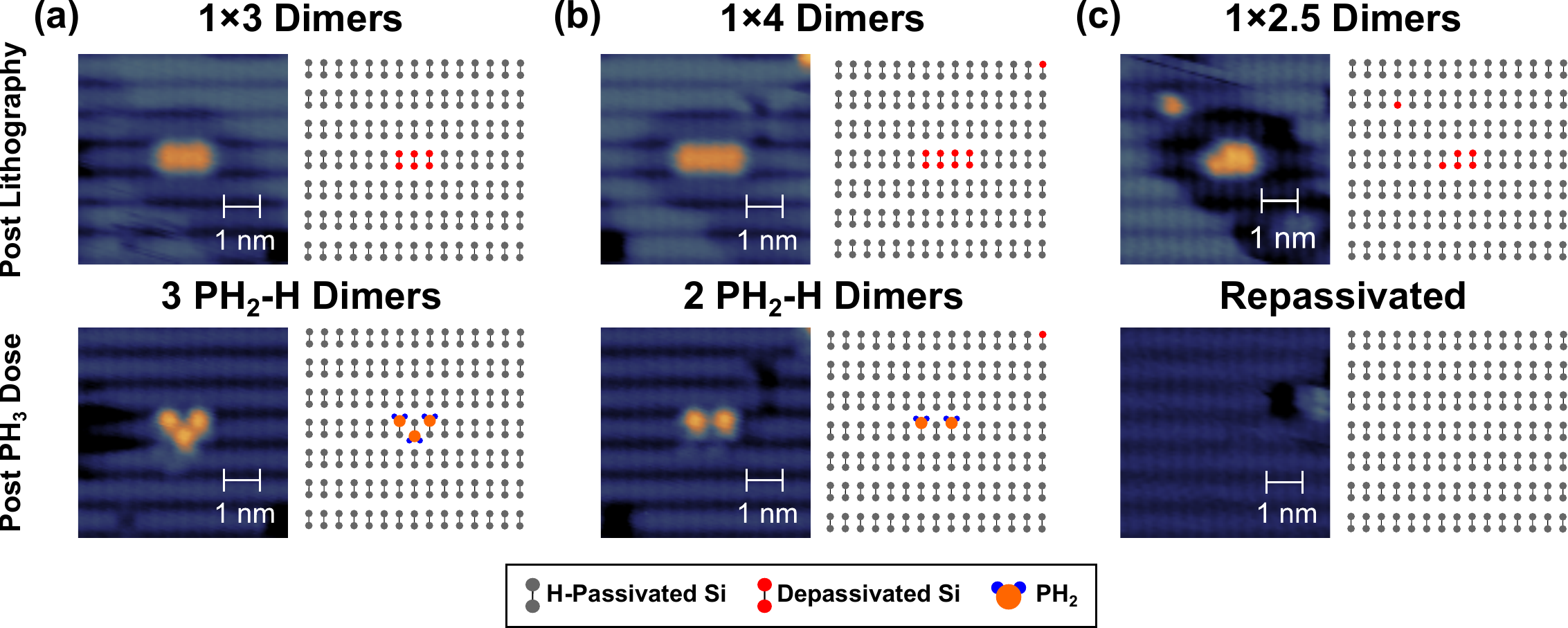}
    \caption{Illustrative PH$_3$ Dosing Outcomes for Lithographic Templates}
    \label{fig:dose-outcomes}
\end{figure}

\begin{figure}[ht]
    \centering
    \includegraphics[width=0.5\columnwidth]{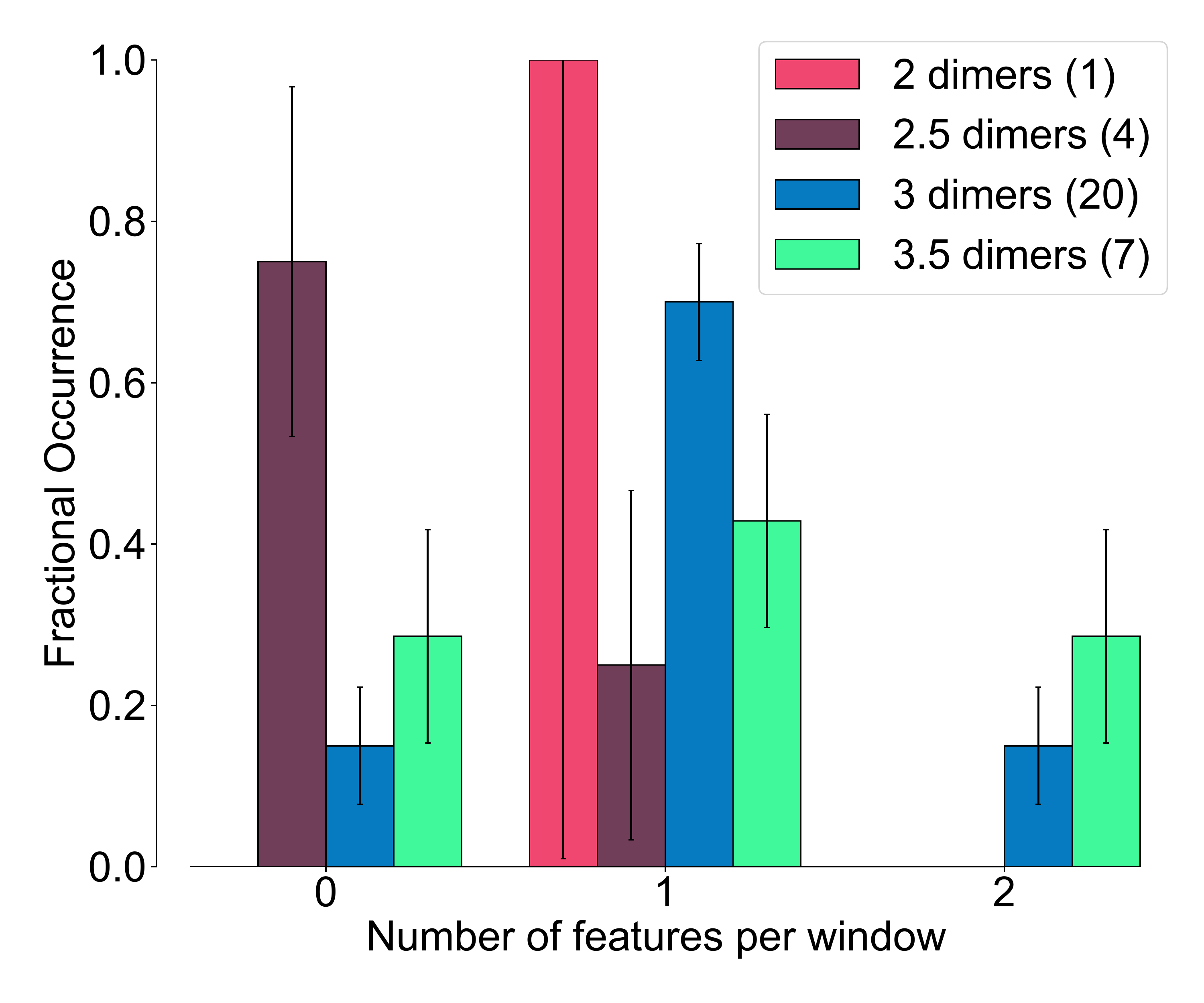}
    \caption{The statistics of how many donors are incorporated expanded to include half dimer windows.}
    \label{fig:half-dimer-stats}
\end{figure}

In the main text, we have compressed results for two ($w=2$) and two and a half dimers ($w=2.5$) into two dimers, and three ($w=3$) and three and a half dimers ($w=3.5$) into three dimers. 
In Fig.~\ref{fig:half-dimer-stats}, we separate these contributions out. 
We find that adding a half dimer to the lithographic window tends to decrease the overall rate of incorporation, although the total sample size, particularly for two dimer wide windows, remains too low to draw significant conclusions. 

\section*{Appendix B: Minesweeper analysis}
\setcounter{equation}{0}
\setcounter{figure}{0}
\renewcommand{\theequation}{B\arabic{equation}}
\renewcommand{\thefigure}{B\arabic{figure}}
\label{app:minesweeper_analysis}

In order to determine $P_I(n|w)$, the pre-dosing and post-incorporation datasets had to be “atomically registered” to each other such that each lattice site corresponds to the same pixel(s) in both datasets. This was accomplished by analyzing and transforming data sets with OpenCV followed by labeling with a custom labeling graphical user interface as detailed in the following.

The first step is to ensure that the pre-dosing to post-incorporation data sets had the same pixel-to-atomic site scaling. This was accomplished by finding the lattice periodicity via Fourier transform and scaling the post-incorporation data to match the scaling of the pre-dosing data sets. We then used the lattice defects as alignment markers between the two data sets by thresholding both data sets and nulling out all data above the approximate height of the defects. A correlation function was applied to the datasets to find the best overlap. Occasionally this produced an insufficient matching and an affine transform was produced by matching three specific defects across the images. Both methods produced a transform function that effectively “atomically registered” the pre-dosing and post-incorporation datasets to one another.

After the images had been “atomically registered”, the datasets were segmented into small regions around the depassivation window and corresponding post-incorporation area. These sub-images were loaded into a custom labeling graphical user interface where the authors shifted each sub-image to adjust the image so that each lattice site corresponded to an 8x8 set of pixels. Each lattice site is labeled as XXX, YYY through analysis of the contrast relatively to the neighboring lattice sites in the STM topographical image. In order to account of uncertainty in both the registration and feature height assignment procedures, both Si ad atom and Si:P heterodimer features within a lithographic window are labeled as well as Si:P heterodimer features found within a single $2\times1$ dimer away from a given window. Such a Si:P heterodimer is included in the analysis only if neither a Si ad atom or Si:P heterodimer is found within the lithographic window, and is considered conclusive evidence of P incorporation as Si:P heterodimers can only result from P incorporation. Several features which would correspond to Si ad atoms were rarely found to be located close to a patterned windows (within several $2\times1$ dimers) where no features were assigned, however as no Si:P heterodimers were found within those respective windows or surrounding them, we do not conclude that these ad atoms indicate the presence of P incorporation. After labeling, each labeled site is located within the overall post-incorporation scan, with the corresponding feature height measured relative to the surrounding interdimer H-resist. In both the pre-dosing and post-incorporation scans, care was taken to avoid incorrect height assignments resulting from “halo” that surrounds dangling bonds and other absorbates on the H-terminated Si$(100)$-$2\times$1 surface while using empty-state imaging conditions \cite{schofield2013quantum}.

%\\begin{figure}[h]
%\    \centering
%\    %\includegraphics{}
%\   \caption{\andrew{A figure that outlines the process and results of the minesweeper analysis. I don't know whether there is two columns of material, but as the other figures are large, then maybe we plan on making this %large too?}}
 %   \label{fig:minesweeper}
%\end{figure}
\section*{Appendix C: Incorporation model}
\setcounter{equation}{0}
\setcounter{figure}{0}
\renewcommand{\theequation}{C\arabic{equation}}
\renewcommand{\thefigure}{C\arabic{figure}}
\label{app:incorporation_model}

%\quinn{Going to insert quick write up of favoring up/down dimers and concerted reaction}
\subsection{Novel Density Functional Theory Configurations}
Our Kinetic Monte Carlo model uses reactions and rates that are aggregated from results in the extensive literature on PH$_3$ dissociation on this particular surface~\cite{warschkow2005phosphine,schofield2006phosphine,wilson2006thermal,radny2007single,bennett2009diffusion,warschkow2016reaction}. 
In addition to the phosphine configurations and reactions, we add an initial adsorption condition and a concerted reaction.

Our initial condition (which is implicit in Ref.~\onlinecite{warschkow2016reaction}, but we are explicitly stating here), relates to the up--down tilt of silicon dimers a bare Si$(100)$-$2\times$1. 
We predict phosphine to have an adsorption energy of --0.91 eV at the lower side of the silicon dimer as opposed to an adsorption energy of --0.56 eV on the higher side. 
In our model, we therefore assume that a phosphine molecule will preferentially adsorb on the lower end of the silicon. 

We also discover an additional concerted reaction (labeled as B2 to C1 in Fig.~\ref{fig:KMC_results}b within the main text). 
In this reaction, a PH$_2$ in the favorable lowered dimer position loses a hydrogen atom to a nearby dimer and moves to a bridging PH position at the same time. This reaction is exothermic with a thermodynamic gain of --1.03 eV and a barrier of 0.96 eV. 
This contrasts with the similar reaction pathway (B3 to C1) from Warschkow \textit{et al.} as the PH$_2$ molecule starts in the thermodynamically favored lower end of the silicon dimer (B2) instead of the higher end (B3), which requires an endothermic reaction to move the PH2 from the favored lowered end (B2) to the higher end (B3) with a thermodynamic cost of 0.24 eV and a barrier of 0.92 eV. 
(While B3 to C1 is technically a two step reaction within the Warschkow \textit{et al.} framework, since the second step is thermodynamically downhill and barrierless, we have consolidated it as one reaction within this work.)

Electronic structure theory calculations for these additions are performed with the Gaussian-basis, local-orbital pseudopotential code {\sc SeqQuest}.\cite{schultzSeqQuest} 
We use a norm-conserving Perdew-Burke-Ernzerhof (PBE) functional.\cite{perdew1996generalized} 
A seven layer thick 4$\times$4 supercell slab of the Si$(100)$-$2\times$1 is constructed and the dangling bonds at the unreconstructed bottom surface were terminated with selenium atoms, as these were found to minimize strain on the slab. 
We use a 20 {\AA} vacuum region to ensure the surfaces at either end of the slab are sufficiently isolated from each other. 
We sample the Brillioun zone with a 2$\times$2$\times$1 grid. 
To achieve total energies relaxed within $\sim$0.01 eV of the structural minimum, atomic positions are relaxed until the forces are reduced to less than 0.0003 Ry/Bohr.

The barriers described above are then used as inputs in an Arrhenius model~\cite{Arrhenius1889rates} for the transition rate $\Gamma = A \exp{\Delta/k_{\rm B}T}$, where $A$ is the attempt frequency, $\Delta$ is the reaction barrier from DFT, $k_{\rm B}$ is Boltzmann's constant, and $T$ is the temperature.
We consider $A=\SI{1e12}{\per \second}$ for all reactions, which is the order of magnitude calculated for these reactions by Warschkow \textit{et al.}~\cite{warschkow2016reaction}.

We calculate the effusive flow rate of molecules landing on any particular silicon dimer as $\Phi_{effusion} =PA/\sqrt{2\pi m k_{\rm B}T}$, where $P$ is the pressure of the incoming precursor gas, $A$ is the area of impingement, taken here as a single silicon dimer, $m$ is the mass of the precursor gas, $k_{\rm B}$ is the Boltzmann constant, and $T$ is the temperature.

We repeat each KMC calculation 200 times to obtain a meaningful statistical sampling of likely outcomes and report the average outcomes, along with standard deviations as applicable. 
For ease of reproducibility, we have placed our Kinetic Monte Carlo code on GitHub~\cite{kmccode}.
We calculate the standard error by assuming a binomial distribution of measured counts and using the standard error based on sample size. 

\subsection{Low-Temperature Dosing Results}

\begin{figure}[ht]
    \centering
    \includegraphics[width=\columnwidth]{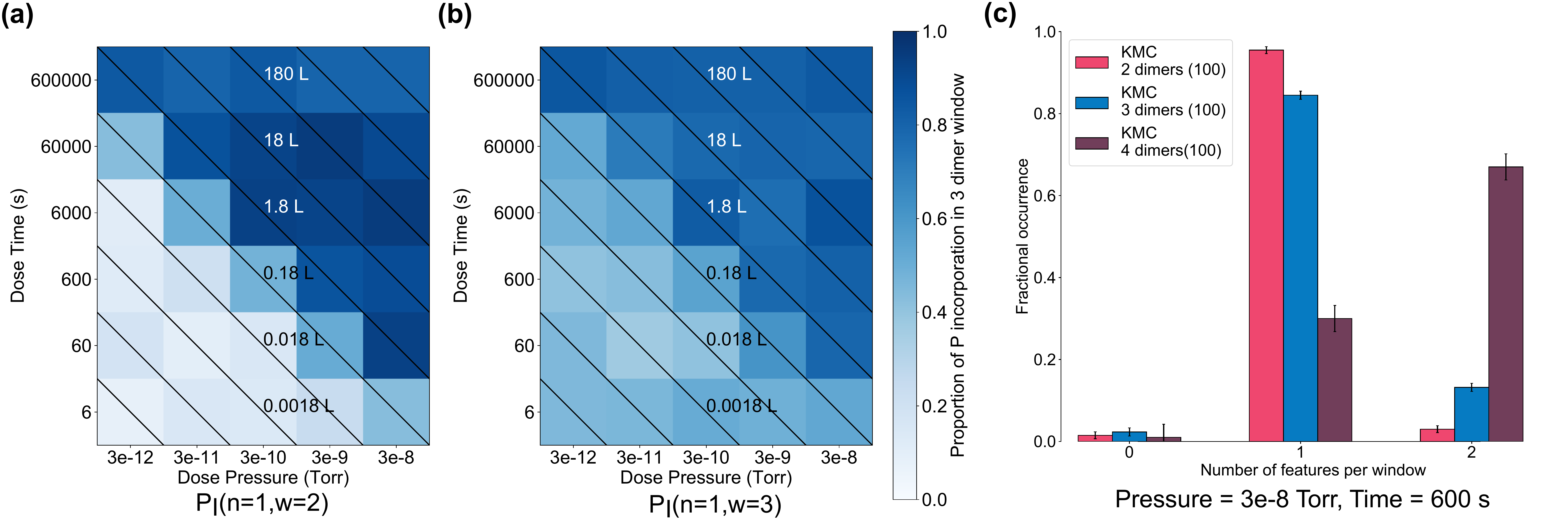}
    \caption{Incorporation rates of a low-temperature dosing scheme, where the dimer windows are initial dosed at 100 K, and then annealed at \SI{500}{\celsius}. (a) The incorporation rate for dimer windows of width $w=2$ as a function of both dosing pressure and temperature. (b) The incorporation rate for dimer windows of width $w=3$ as a function of both dosing pressure and temperature. (c) The relative frequencies of differing numbers of incorporations as a function of dimer width with a dose pressure of \SI{3e-8}{Torr} and a dose time of 600 s. Notably, a significant portion of three dimer window exposures result in two incorporation events.}
    \label{fig:low-T-dosing}
\end{figure}

We also use our kinetic Monte Carlo model to examine low-temperature ( $<$ 100 K) dosing techniques to achieve saturation before annealing such as in a recent patent.\cite{simmons_keizer_2019}
We simulate a process where the initial dosing step is done at a temperature of 100 K, and subsequent annealing at a temperature of \SI{500}{\celsius} for 15 s.
As shown in Fig.~\ref{fig:low-T-dosing} in saturated dosing conditions (\textit{i.e.} $>$ 1 L coverage), near-deterministic doping can be achieved. 

Dosing at low temperatures allows saturation coverage of the exposed window to occur. At 100 K, enough energy is provided for an initially adsorbed phosphine to shed one hydrogen, but not for further reactions.
The subsequent anneal at \SI{500}{\celsius} provides enough energy for the phosphine fragments to desorb in a reasonable time frame. 
This provides enough room for the remaining phosphine fragments to fully dissociate into bridging PH fragments. 
For a 2 dimer wide window, this process is nearly deterministic with a predicted incorporation rate of $P_I(n=1|w=2)=96\%$ for a dose pressure of \SI{3e-8}{Torr} and a dose time of 600 s, as shown in Fig.~\ref{fig:low-T-dosing}c.
However, for 3 dimer window, this process leads to two incorporations with a non-negligible probability; $P_I(n=2|w=3)=13\%$ for a dose pressure of \SI{3e-8}{Torr} and a dose time of 600 s.

While this process can lead to near deterministic incorporation, it is also highly likely to lead to hydrogen desorption in the surrounding resist. 
This is due to the barriers for H$_2$ desorption and PH$_2$ + H desorption being essentially the same.
After a low-temperature dose, the entire window will be saturated with PH$_2$ + H fragments and not full PH$_3$ molecules, making the PH$_2$ desorption barrier the relevant rate limiting step for full dissociation.
Desorption of PH$_2$ + H has a barrier of approximately 2.0 eV, while H$_2$ desorption has a barrier of 2.1 eV.\cite{wilson2006thermal}
Any anneal sufficiently hot to desorb PH$_2$ in a reasonable time frame would therefore almost certainly also desorb hydrogen from the surrounding silicon and destroy the integrity of the surrounding resist. 
While the small difference  in barriers between PH$_2$ + H and H$_2$ desorption can theoretically be exploited to find an anneal schedule that maximizes PH$_2$ + H dissociation while leaving H$_2$, we find that this would lead to non-deterministic incorporation.
Using a 100 K dose with a 10 minute \SI{400}{\celsius} anneal had the best effect for maximally encouraging PH$_2$ + H dissociation with minimal hydrogen resist degradation, but the maximum incorporation rate was only $P_I(n=1|w=3)=79\%$.

This analysis predicts that saturation dosing at low temperature and then high temperature annealing techniques are a viable route to deterministic incorporation, albeit with the potentially significant drawback of losing hydrogen resist integrity and thus introducing more uncertainty into the precise location of incorporation.
We are careful to note the obvious fact that experiment is the ultimate arbiter of whether the resist integrity is degraded in practice.
Combined with the findings in the main text, this indicates that the experimentally stochastic incorporation rates may, in fact, be an artifact of room temperature dosing.
Other regimes of both low and moderate temperature dosing with the correct combination of dose pressure and time are predicted to be more feasible routes toward atomic precision deterministic single donor doping.

\section*{Appendix D: Extended Fermi-Hubbard Hamiltonian and Transport Modeling}
\setcounter{equation}{0}
\setcounter{figure}{0}
\renewcommand{\theequation}{D\arabic{equation}}
\renewcommand{\thefigure}{D\arabic{figure}}
\label{app:extended_fermi-hubbard_parameters}

Similar to other authors~\cite{le2017extended,dusko2018adequacy}, we describe our donor chains with the extended Fermi-Hubbard Hamiltonian, which takes the form
\begin{equation}
    \hat{H} = \sum \limits_{j,\sigma} \epsilon_{j,\sigma} \hat{n}_{j,\sigma} + \sum \limits_{<j,k>,\sigma} t_{j,k} \left( \hat{c}^{\dagger}_{j,\sigma} \hat{c}_{k,\sigma} + \hat{c}^{\dagger}_{k,\sigma} \hat{c}_{j,\sigma} \right) + U \sum \limits_{j} \hat{n}_{j,\uparrow} \hat{n}_{j,\downarrow} + \sum \limits_{j,k} V_{j,k} \hat{\rho}_{j} \hat{\rho}_{k}  \label{eq:extended_fh_hamiltonian}.
\end{equation}
The four sums comprising this Hamiltonian describe the following physics:
\begin{enumerate}
    \item The energy ($\epsilon_{j,\sigma}$) of a single-particle orbital localized on site $j$ with spin $\sigma$. 
    This includes the effects of externally applied electric fields. 
    The particle number operator for site $j$ and spin $\sigma$ is $\hat{n}_{j,\sigma}=\hat{c}^{\dagger}_{j,\sigma} \hat{c}_{j,\sigma}$, where $\hat{c}^{\dagger}_{j,\sigma}$ ($\hat{c}_{j,\sigma}$) is the associated creation (annihilation) operator associated with that fermionic spin-orbital.
    \item The spin-independent tunnel coupling ($t_{j,k}$) between sites $j$ and $k$. 
    The brackets in the summation indicate that it is restricted to nearest-neighbors, though we note that a more detailed examination of this assumption for arrays with short pitches will be the topic of future work.
    \item The Coulomb repulsion ($U$) associated with double occupation of site $j$ by electrons with opposite spins.
    \item The Coulomb repulsion ($V_{j,k}$) between electrons occupying sites $j$ and $k$. 
    The density operator associated with site $j$, $\hat{\rho}_j=\hat{n}_{j,\uparrow}+\hat{n}_{j,\downarrow}$. 
    We note that the conventional version of this model is limited to nearest-neighbor intersite repulsion.
\end{enumerate}

We rely on an implementation of multi-valley effective mass theory (MV-EMT)~\cite{gamble2015multivalley} first developed at Sandia National Laboratories for the parameters of this model.
This particular approach to MV-EMT due to Shindo and Nara~\cite{shindo1976effective} goes beyond the more common Kohn-Luttinger theory~\cite{kohn1955theory} and includes a non-perturbative and consistent treatment of both the central-cell correction and valley-orbit coupling that captures the fine structure of the valley-orbit split $1s$-like states for a single donor.
Recent extensions of this implementation of MV-EMT that account for electron-electron interactions in the modeling of exchange-coupled donors indicate that this theory can be adjusted to capture the charging energy of a single doubly-occupied donor, as well~\cite{joecker2020full}.
The $t_{j,k}$ parameters are taken from an exhaustive set of tunnel couplings between individual pairs of donors in bulk silicon that were made available in the arXiv version of Ref.~\onlinecite{gamble2015multivalley}~\cite{gamble2014multivalley}.
The onsite repulsion is $U=\SI{43.8}{\milli \eV}$ to reproduce the charging energy of a single bulk-like $D^{-}$ state.
$V$ is simply the static screened Coulomb interaction between charge centers localized on donor sites with $\epsilon_{Si}=11.7$.

We expect the accuracy of this particular parameterization will be lacking relative to an experiment for a number of reasons.
The tunnel couplings were computed between pairs of bulk-like donors, the $U$ term does not account for the fact that the doubly occupied orbitals will have a different character in a chain with a short pitch (i.e., we expect $U$ to be lower), and overall we expect that the single-particle orbitals will look quite different from our base MV-EMT.
All of these issues are compounded by the fact that we are studying non-uniform chains in which some sites have failed to incorporate.
For the purposes of the qualitative analysis in this paper, we argue that these limitations are acceptable and that experimental design and interpretation will require a more careful parameterization.
We note that ongoing work is aimed at developing such a parameterization using an mesh-based MV-EMT capability that makes use of a discontinuous Galerkin discretization of the Shindo-Nara equations and device-specific self-consistent electrostatics.
While donor chains are particularly challenging to model with this level of detail, results from this solver (adapted to the Luttinger-Kohn Hamiltonian) have been presented for the analysis of relatively simple Ge hole quantum dots in Ref.~\cite{miller2021effective}.

To simulate bias spectroscopy, we use the Meir-Wingreen formula~\cite{meir1992landauer} to compute the current through a donor chain modeled as an instance of the extended Fermi-Hubbard Hamiltonian in Eq.~\ref{eq:extended_fh_hamiltonian}.
We use exact diagonalization to compute the necessary Green's functions, which is computationally tractable for this simplistic model in which each site is comprised of two spin orbitals.
A two-point finite difference formula is used to compute the differential conductance from the current, computed at 501 points for each instance.

\begin{figure}[ht]
    \centering
    \includegraphics{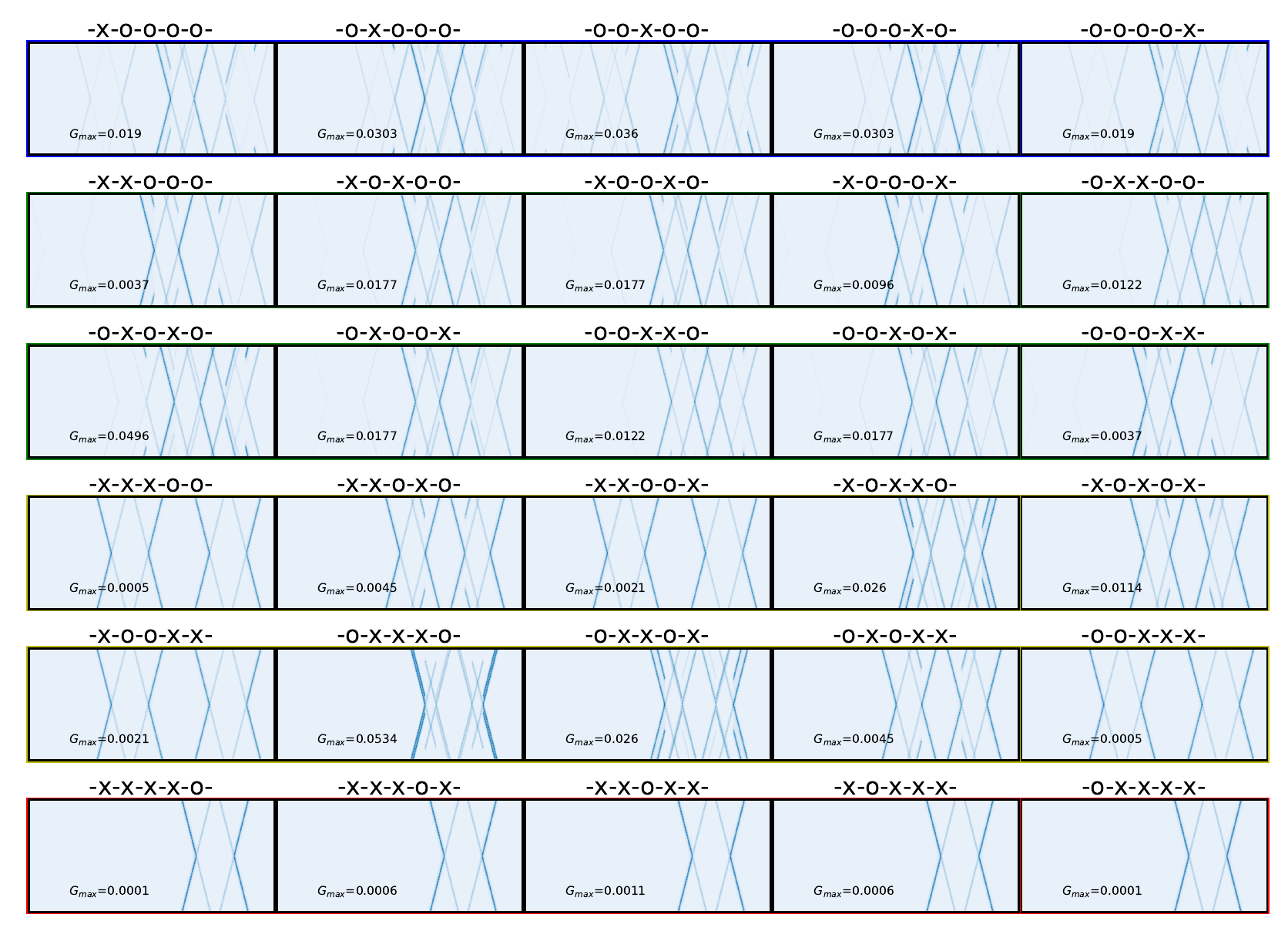}
    \caption{Exhaustive set of bias spectra for a 5-donor chain in which $n \leq 4$ sites are missing.
    Here the spacing between donors was chosen to be \SI{2.72}{\nano \meter} along [100].
    The x axes are all identical corresponding to varying $\epsilon_{j,\sigma}$ from \SI{-0.5}{\eV} to \SI{0.1}{\eV}.
    The y axes are all identical corresponding to varying the source-drain bias from \SI{-0.07}{\eV} to \SI{0.07}{\eV}.
    }
    \label{fig:ensemble_statistics_2.5nm_100}
\end{figure}

\begin{figure}[ht]
    \centering
    \includegraphics{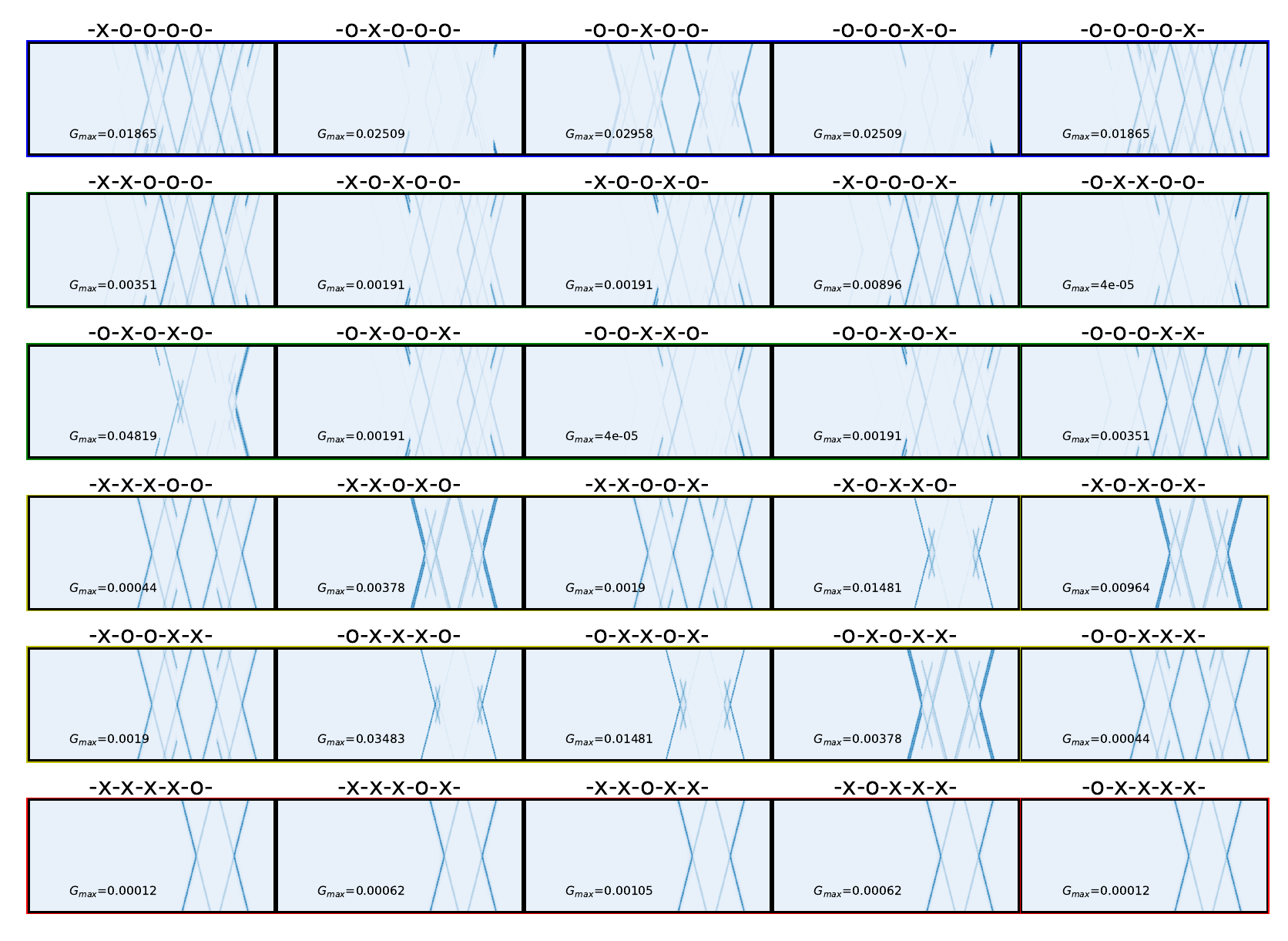}
    \caption{Exhaustive set of bias spectra for a 5-donor chain in which $n \leq 4$ sites are missing.
    Here the spacing between donors was chosen to be \SI{5.43}{\nano \meter} along [100].
    The x axes are all identical corresponding to varying $\epsilon_{j,\sigma}$ from \SI{-0.5}{\eV} to \SI{0.1}{\eV}.
    The y axes are all identical corresponding to varying the source-drain bias from \SI{-0.07}{\eV} to \SI{0.07}{\eV}.}
    \label{fig:ensemble_statistics_5nm_100}
\end{figure}

\begin{figure}[ht]
    \centering
    \includegraphics{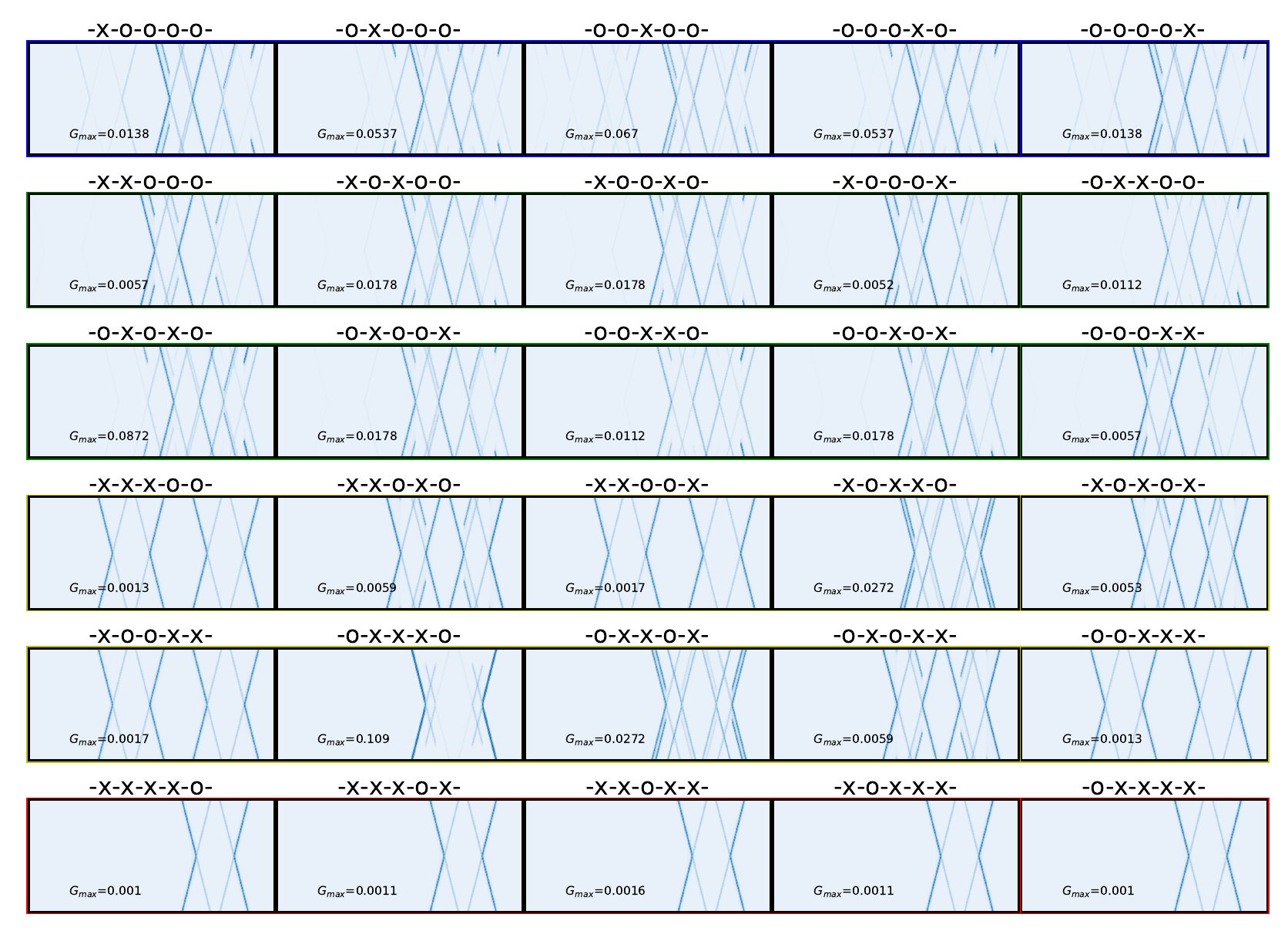}
    \caption{Exhaustive set of bias spectra for a 5-donor chain in which $n \leq 4$ sites are missing.
    Here the spacing between donors was chosen to be \SI{2.69}{\nano \meter} along [110].
    The x axes are all identical corresponding to varying $\epsilon_{j,\sigma}$ from \SI{-0.5}{\eV} to \SI{0.1}{\eV}.
    The y axes are all identical corresponding to varying the source-drain bias from \SI{-0.07}{\eV} to \SI{0.07}{\eV}.}
    \label{fig:ensemble_statistics_2.5nm_110}
\end{figure}

\begin{figure}[ht]
    \centering
    \includegraphics{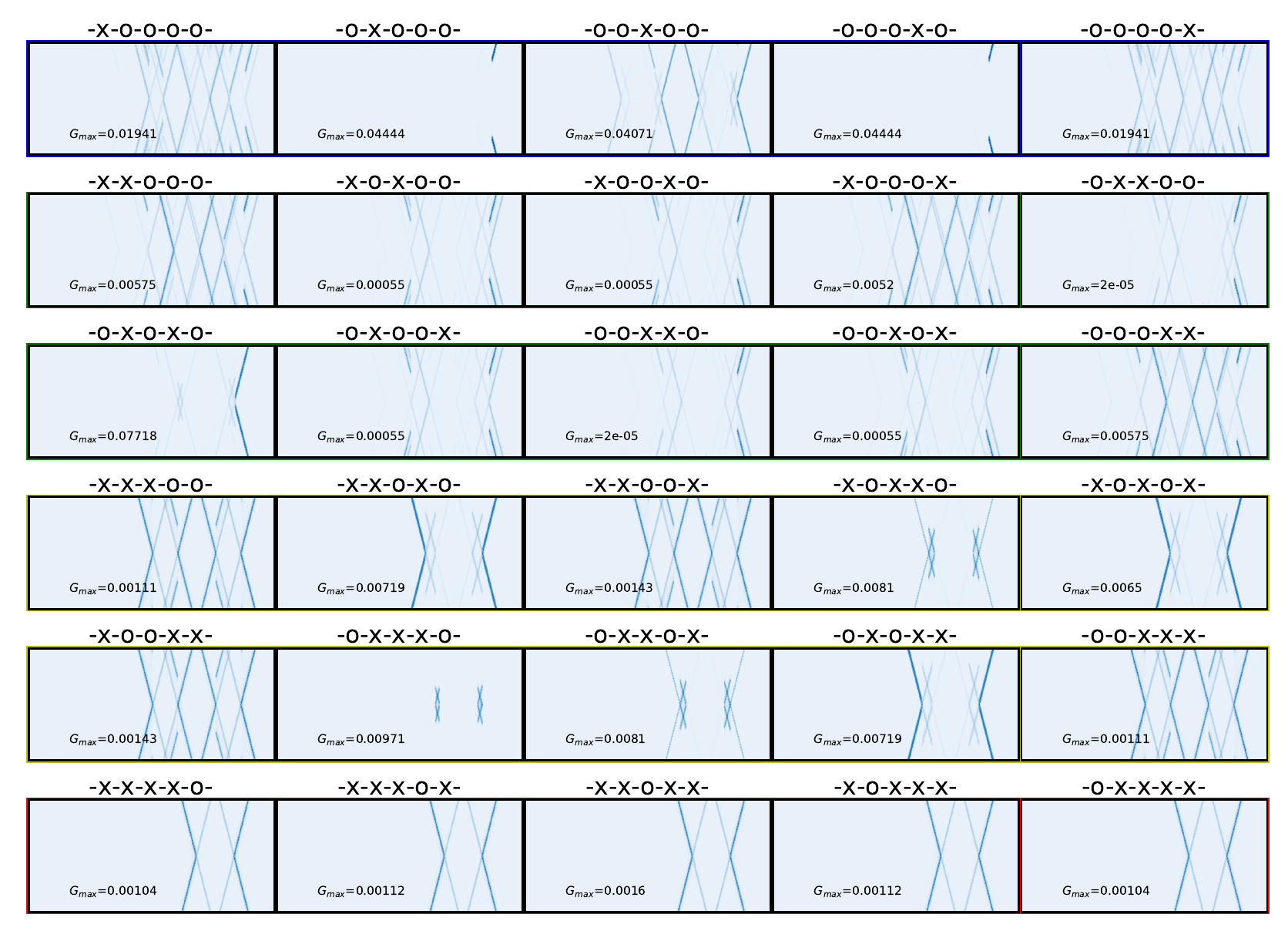}
    \caption{Exhaustive set of bias spectra for a 5-donor chain in which $n \leq 4$ sites are missing.
    Here the spacing between donors was chosen to be \SI{5.38}{\nano \meter} along [110].
    The x axes are all identical corresponding to varying $\epsilon_{j,\sigma}$ from \SI{-0.5}{\eV} to \SI{0.1}{\eV}.
    The y axes are all identical corresponding to varying the source-drain bias from \SI{-0.07}{\eV} to \SI{0.07}{\eV}.}
    \label{fig:ensemble_statistics_5nm_110}
\end{figure}

We present simulated bias spectra for the exhaustive set of one-dimensional chains with two pitches, corresponding to $\approx$ \SI{2.5}{\nano \meter} and $\approx$ \SI{5}{\nano \meter} spacings along [100] in Fig.~\ref{fig:ensemble_statistics_2.5nm_100} and~\ref{fig:ensemble_statistics_5nm_100} (respectively) and [110] in Fig.~\ref{fig:ensemble_statistics_2.5nm_110} and Fig.~\ref{fig:ensemble_statistics_5nm_110}.
The spacing and length total length of the chain are chosen to be sufficiently short that it is reasonable to expect that the tunneling current can be measured even in the absence of missing sites. 
We note that our calculations do not include current through scattering states to clearly demarcate the transport through the eigenstates of the chain and thus our Coulomb diamonds do not have signatures of transport ``above'' or ``below'' them, as would be expected in a real experiment.

In each of the following figures, the individual rows correspond to a particular number of single-donor sites missing.
For a 5-donor chain, there are thus 20 distinct non-trivial instances (i.e., 5 instances with one donor missing, 10 with two, 10 with three, and 5 with four).
These are indicated on top of each bias spectrum with ``x'' indicating a missing site and ``o'' indicating a donor.
The scale of the color bars is allowed to vary to capture the relevant details in each plot and only the maximum value of the differential conductance is specified in each panel.

\end{document}